\documentclass{emulateapj}
\usepackage{amsmath}
\usepackage{color}
\usepackage{soul}

\newcommand{\funit}{ph~cm$^{-2}$~s$^{-1}$}

\slugcomment{Accepted for publication in {\sc the astrophysical journal}: 2016 July 12}

\shorttitle{{\em Fermi}-LAT Observations of the 2014 May--July outburst from 3C~454.3}
\shortauthors{Britto, Bottacini, Lott, Razzaque \& Buson}

\begin{document}

\title{{\em Fermi}-LAT Observations of the 2014 May--July outburst from 3C~454.3}

\author{Richard J. Britto\altaffilmark{1,2}, Eugenio Bottacini\altaffilmark{3}, Beno\^it Lott\altaffilmark{4}, Soebur Razzaque\altaffilmark{1} and Sara Buson\altaffilmark{5,6,7,8}}
\altaffiltext{1}{Department of Physics, University of Johannesburg, PO Box 524, Auckland Park 2006, South Africa, \emph{srazzaque@uj.ac.za}}
\altaffiltext{2}{Current affiliation: Department of Physics, University of the Free State, PO Box 339, Bloemfontein 9300, South Africa, \emph{brittor@ufs.ac.za, dr.richard.britto@gmail.com}}
\altaffiltext{3}{W. W. Hansen Experimental Physics Laboratory, Stanford University, Stanford, CA 94305, USA, \em{eugenio.bottacini@stanford.edu}}
\altaffiltext{4}{Univ. Bordeaux, Centre d'\'Etudes Nucl\'eaires de Bordeaux-Gradignan, UMR 5797, CNRS/IN2P3, 33175 Gradignan, France, \emph{lott@cenbg.in2p3.fr}}
\altaffiltext{5}{Istituto Nazionale di Fisica Nucleare, Sezione di Padova, I-34131, Padova, Italy}
\altaffiltext{6}{Dipartimento di Fisica ``G. Galilei'', Universit\`a di Padova, I-34131, Padova, Italy}
\altaffiltext{7}{Current affiliation: Astrophysics Science Division, NASA Goddard Space Flight Center, Greenbelt, MD 20771 USA, \emph{sara.buson@nasa.gov}}
\altaffiltext{8}{Current affiliation: University of Maryland Baltimore County/CRESST, Baltimore, MD 21250, USA}

\begin{abstract}
  A prominent outburst of the flat spectrum radio quasar
  3C~454.3 was observed in 2014 June with the {\em Fermi} Large
  Area Telescope. This outburst was characterized by a three-stage
  light-curve pattern---plateau, flare and post-flare---that occurred
  from 2014 May to July, in a similar pattern as observed
  during the exceptional outburst in 2010 November.
  The highest flux of the outburst
  reported in this paper occurred during 2014 June 7--29, showing a
  multiple-peak structure in the light-curves. The average flux in
  these 22 days was found to be $F[E > 100~\mathrm{MeV}] = (7.2 \pm 0.2) \times 10^{-6}$~\funit,
  with a spectral index, for a simple power law, of
  $\Gamma = 2.04 \pm 0.01$. That made this outburst the first
  $\gamma$-ray high state of 3C~454.3 ever to be detected by {\em Fermi} with
  such a hard spectrum over several days. The highest
  flux was recorded on 2014 June 15, in a 3 hr bin, at MJD 56823.5625,
  at a level of $F[E > 100~\mathrm{MeV}] = (17.6 \pm 1.9) \times
  10^{-6}$~\funit. The rise time of one of the short subflares was
  found to be $T_r= 1200 \pm 700$~s at MJD = 56827, when the flux increased
  from 4 to 12 $\times 10^{-6}$~\funit. Several photons above 20 GeV
  were collected during this outburst, including one at 45 GeV on MJD
  56827, constraining the $\gamma$-ray emission
  region to be located close to the outer boundary of the broad-line
  region, leading to fast flux variability.
\end{abstract}


\section{Introduction}

The flat spectrum radio quasar (FSRQ) 3C~454.3 (also cataloged as PKS 2251+158;
$R.~A.=22^h~53^m~57^s.748$, $decl.=+16^\circ~08'~53''.56$~(2000); redshift z = 0.859 \citep{1967ApJ...147..837L})
is a well-known active galactic nucleus (AGN) that shows very
bright flaring activity across the electromagnetic waveband. The
source was detected above 100 MeV by the {\em Energetic Gamma Ray Experiment Telescope} (EGRET) on board the
{\em Compton Gamma Ray Observatory} in 1999
\citep{1999ApJS..123...79H}.
Flares had been reported by the {\em AGILE} space telescope since 2007
\citep{Vercellone1,Vercellone2} and by the {\em Fermi} Large Area Telescope
({\em Fermi}-LAT) since 2008 \citep{Abdo2009}. Four historical flaring episodes made 3C~454.3 the
brightest $\gamma$-ray source ever recorded in the sky, apart from
$\gamma$-ray bursts, and were studied in detail using {\em Fermi}-LAT
data. These outbursts occurred in 2009 December
\citep{Striani,Ackermann2010}, 2010 April \citep{Ackermann2010},
2010 November \citep{Abdo2011} and 2014 June (this work). The daily
recorded flux levels detected by {\em Fermi}-LAT, $F[E > 100~\mathrm{MeV}]$
($F_{100}$, in units of 10$^{-6}$ photons~cm$^{-2}$~s$^{-1}$), reached
$F_{100} = 22 \pm 1$ in 2009 December and $F_{100} \simeq 16$ in 2010 April
\citep{Striani,Ackermann2010}\footnote{When used as a unit, ``photon'' is hereafter abbreviated as ``ph''}.
In 2010 November, the source
displayed sustained activity at a level of $F_{100} \simeq 10$ for
several days, and then showed a fast rise to $F_{100} \simeq 55$, before
reaching the historic record level of $F_{100} \simeq 80$ in a 6 hr bin, on 2010 November 20 \citep{Abdo2011}.

The 2009--2010 outbursts of 3C~454.3 exhibit similar patterns including three phases,
which were more clearly observable in the light-curve of the
2010 September--December outburst \citep{Abdo2011}. First, there is a \emph{plateau} phase
at the beginning where the flux rapidly increases and remains steady for a few
days at a relatively high state, followed by the main \emph{flare} (that
usually exhibits several distinct peaks). A \emph{post-flare} phase is then
visible, where the flux is more or less fluctuating at a lower level than the flare,
but generally at a higher level than the plateau. This three-phase pattern was first identified and reported in \citet{Ackermann2010} and \citet{Abdo2011}, and then quantitatively described in \citet{2013ApJ...773..147J}, for the three 2009--2010 outbursts.
It is also common to present data of a quiescent phase (referred to as {\em pre-flare}), preceding the
plateau, for comparative studies with the following high-state activity.

A significant hardening of the photon index $\Gamma$ was reported
during the 2010 November outburst \citep{Abdo2011}. For $E > 100$ MeV
and a power-law (PL) model, $\Gamma \simeq 2$ was measured during this
flare in 6 hr bins, compared to $\Gamma \simeq 2.5$ in the quiescent
regime. An even harder index has been reported during the short flare
of 2013 September 23--25, when $\Gamma\simeq 1.82\pm 0.06$ was
measured up to 40 GeV on average during those two days \citep{Pacciani}.

Like many other bright FSRQs, 3C~454.3 shows a spectral break at GeV energies, which was found to be
almost independent of its flux level \citep{Abdo2010}. A change in the PL photon
index was reported to be significantly more than one unit at energies
around 2 GeV \citep{Abdo2009,Ackermann2010,Abdo2011}.  As this
spectral feature was not a theoretical prediction and was observed
only with {\em Fermi}-LAT from 2008, different models were proposed only in
recent years to account for it. Indeed, this
broken spectrum was found inconsistent with a $\gamma$-ray photon
distribution produced by inverse-Compton (IC) scattering due to a
cooling electron distribution. \citet{Finke1} proposed a combination
of the Compton-scattered disk and broad-line region (BLR) radiation to explain the
spectral break and also fit the quasi-simultaneous radio, optical/UV,
X-ray and $\gamma$-ray spectral energy distributions (SEDs), using data
from the 2008 August flare. Also, \citet{Cerruti} proposed that the
break feature can be explained by a log-parabolic distribution of
non-thermal electrons that up-scatter photons in the BLR, using data from both the 2008 August and the
2010 November flares. By studying the effect of a continuous time-dependent particle injection of electrons into the jet, \citet{Hunger} proposed a phenomenological model-independent approach which was applied to the 2010 November flare. They found that the combination of the Compton-scattered disk and BLR radiation, and the Compton-scattered BLR radiation only with an intrinsic break in the ambient particle distribution, are both viable scenarios, for specific injection parameters.
More recently, \citet{Kohler} studied short bright
$\gamma$-ray flares of several blazars, including 3C~454.3, and reported variable spectral features that 
significantly depart from the average spectrum that is obtained over several months/years.
They suggest that the average {\em Fermi}-LAT spectrum does not reflect a specific particle acceleration scenario, since it appears to be the superposition of several short-lived components,
each one exhibiting a different spectral curvature.
The observed SEDs would rather reflect macroscopic turbulence in relativistic jets
with emitting regions having a narrow energy distribution of emitting particles.

Exotic scenarios such as mixing of photons with axion-like
particles in the large-scale jet has also been found to fit the
spectra of 3C~454.3 in its 2010 November outburst
\citep{2013JCAP...11..023M}.

Another important issue that is investigated in the study of FSRQs, in
connection with the GeV spectral break, is the location of the $\gamma$-ray
emission region. Observations at GeV energies could constrain
the location of the emission region to be within or beyond
the BLR \citep{Dotson,Orienti,Nalewajko}. More specifically, for each
flare, different attempts were made to constrain the location of the $\gamma$-ray emitting blob. 
\citet{Fuhrmann} performed radio/$\gamma$-ray correlation studies of 54 bright {\em Fermi} blazars, including 3C~454.3, on a 3.5-year data sample.
Based on a discrete cross-correlation function analysis method, they measured time lags between light-curves at different frequencies and the $\gamma$-ray light-curve, constraining the $\gamma$-ray emitting region to lie between 0.8 and 1.6 pc from the super-massive black hole (SMBH) for 3C~454.3. For the same source, the 3 mm $\tau=1$ optical depth surface is found to lie between 2 and 3 pc from the SMBH, which is higher than the typical bulk BLR radius of $\sim$0.2 pc for this source. This would suggest that $\gamma$ rays are produced beyond the canonical BLR and that some material of the BLR also lies at these larger distances.
However, it is also interesting to study the steady $\gamma$-ray emission over a long period of time, as performed by \citet{Poutanen}. These authors have presented studies from 180 days of {\em Fermi}-LAT data to model the spectral breaks, using
the pair production mechanism from $\gamma$ rays of the jet interacting
with UV photons of the BLR. In \citet{Stern2011}, they propose a model
for 3C~454.3 (data collected in 2008--2011) in which $\gamma$ rays are
produced close to the boundary of the high-ionization part of the BLR,
and move away from the black hole with the increase of the flux. In
their latest study of 1740 days of {\em Fermi}-LAT data from bright blazars,
they propose that a $\sim$5 GeV break is mainly due to He {\sc ii} Ly
continuum photons at 54.4 eV, and that a $\sim$20 GeV break is mainly due to
the H Ly continuum. In the case of 3C~454.3, these would give a
significant constraint on the location of the $\gamma$-ray emission region
to be within the BLR \citep{Stern2014}.

A multiwavelength study of 3C~454.3 data obtained during the period 2005--2008 was performed by \citet{Jorstad}, reporting correlation between optical,
X-ray and $\gamma$-ray variations, and proposing that the $\gamma$-ray
emission is dominated by the external Compton (EC) mechanism. \citet{Pacciani2010}
used the 2009 November--December multiwavelenghth campaign data on the bright flare of 3C~454.3 to model the pre- and post-flare broadband SED with a one-zone synchrotron self-Compton plus EC emission model.  However, they determined that the SED around flare maximum required an additional particle component to achieve a satisfactory fit.
On the other hand, the study of the same 2009 December
flare by \citet{Bonnoli} led them to the conclusion that a simple
one-zone synchrotron + IC emission can model the broadband SED, the IC emission consisting of synchrotron self-Compton in the X-ray band, in addition to EC.

\citet{2013ApJ...763L..36L} studied the flux variability of the Mg {\sc ii} $\lambda$2800 emission line of 3C~454.3 by using optical spectra acquired as part of the ``Ground-based Observational Support of the \emph{Fermi} Gamma-Ray Space Telescope at the University of Arizona'' monitoring program. This shows interesting perspectives for multiwavelength studies during flaring events, and to probe the evolution of the radiation field of the BLR.

In this paper we report on the 2014 May--July outburst of 3C~454.3 as
observed by {\em Fermi}-LAT. Although the flux ($F_{100}$) during this
latest outburst did not reach the level of the 2010 November flare, it
lasted longer and we could separate its different substructures quite
well, compared to the previous flares. This $\gamma$-ray outburst was reported by the \emph{AGILE} team
in \citet{ATEL_AGILE} during its plateau phase, and by the \emph{Fermi}-LAT team in \citet{2014ATel.6236....1B} during the main flare.
We also observed that the flux
remained quite high (very often at $F_{100} > 2$ daily) for several weeks
after the flare. The plateau--flare--post-flare pattern was clearly
observed as well. During this outburst, a spectrum significantly
harder than in 2010 was measured, with most of the flaring phase being observed
with a hard photon spectral index $\Gamma \simeq 2$,
mainly during the MJD 56827.0--56833.5 period.
To the best of the authors' knowledge, this is the first $\gamma$-ray high state of 3C~454.3 ever to be detected
with such a hard PL photon index, except for the 2013 September
flare that was too short to be characterized in detail in its temporal as
well as its spectral structure \citep{Pacciani}.  The tendency of a
``harder-when-brighter'' pattern for 3C~454.3 was clearly visible
during this (2014) outburst, when comparing the spectral shape of the different phases.

We study flux variability and change in spectral properties in the
$\gamma$-ray band at different epochs during the 2014 May--July outburst
of 3C~454.3. Detection of a flux rise time of $\sim$1200 s in one of
the short duration subflares is one of the shortest in the GeV band for
3C~454.3. A 45 GeV photon was detected with high confidence during
this subflare, allowing us to put a lower limit on the jet Lorentz
factor that is the highest to date along with what was estimated by \citet{Jorstad} and \citet{Sikora}.
Another high-energy (HE), 39 GeV, photon was also detected
during this outburst.

X-ray, optical and near infrared light-curves were seen to be well
correlated with the {\em Fermi}-LAT $\gamma$-ray data during this
latest flare, though the radio light-curve shows a global tendency of
continuous increase during the main outburst structure \citep{ATEL_NIR,ATEL_MIRO,ATEL_SMARTS}.
\citet{Tachibana} reported the analysis of optical
versus $\gamma$-ray data in the MJD 56800--56910 period, showing some
evidence of a change in the Doppler factor $\delta$ during the flare.

In Section 2, observations and analysis of {\em Fermi}-LAT data for
the outburst phases are presented, including light-curves, HE ($\gtrsim 10$ GeV) photons and $\gamma$-ray luminosities, fastest
variability studies and SEDs. In section 3, we present the analysis of {\em Swift}-XRT data. Results are presented in Section 4 and a discussion in Section 5. A flat $\Lambda$-CDM cosmology with
$H_0$ = 69.6~km~s$^{-1}$ Mpc$^{-1}$, $\Omega_m$ = 0.286, and
$\Omega_\Lambda$ = 0.714 is used in this paper \citep{Planck}.

\section{{\em Fermi}-LAT observations and analysis} \label{sec:obs_ana}
The {\em Fermi}-LAT is a pair-conversion $\gamma$-ray telescope sensitive to photon energies
greater than 20 MeV with a field of view of about 2.4 sr \citep{Atwood}. The data presented in this paper
were collected during the 2013 September 10--2014 August 21 period (MJD 56545.0 to 56890.0). During this period, in contrast to the nominal scanning mode where the LAT surveys the whole sky every 3 hr, different \emph{target of opportunity} observing modes were in place, in order to increase the exposure of PSR B1259-63. Also, due to an instrument pointing issue, no data were collected on the whole sky during a $\sim$13 hr long period, on 2014 June 26 (MJD 56834.07--56834.65), leading to a gap in the light-curves shown below
(corresponding to the ``peak 6'' phase).

Only photons with energies greater than 100 MeV were considered
in this analysis. In order to avoid contamination from the Earth limb $\gamma$ rays, a selection
of events with zenith angle $\theta_z<90^{\circ}$ was applied.
This analysis was performed with the
standard analysis tool {\em gtlike/pyLikelihood}, which are part of the {\em Fermi} Science Tools
software package (version v10r0p5).\footnote{http://fermi.gsfc.nasa.gov/ssc/data/analysis/software/}
The {P8R2\_SOURCE\_V6} set of instrument response functions (IRFs) was used.
We selected the corresponding source-class events above 100 MeV. Compared to the \emph{Pass 6} analyses of previous outbursts of 3C 454.3 presented in \citet{Ackermann2010} and \citet{Abdo2011}, \emph{Pass 7 REPROCESSED} provided a higher effective area below ~200 MeV and a better angular resolution above a few GeV, in addition to improved models for the diffuse emission components \citep{2012ApJS..203....4A,2013arXiv1304.5456B}. The Pass~8 data representation\footnote{http://www.slac.stanford.edu/exp/glast/groups/canda/lat\_\\Performance.htm} now provides a significant improvement in terms of acceptance and energy resolution, using more in-flight data for a better calibration of the event reconstruction and background rejection (\citet{2013arXiv1303.3514A}, see also Appendix A of \citet{2016ApJ...819..149A}). The increase of $\sim$15\% in the effective area in the 10--50 GeV range---with respect to Pass 7 REPROCESSED---was particularly interesting for the study of the HE photons presented in Section \ref{subsec:HE_L}.

In the analysis presented in this paper, photons were selected in a 10$^{\circ}$ radius region of interest (ROI), centered at the position of 3C~454.3. The isotropic background, including the sum of residual instrumental background and extragalactic
diffuse $\gamma$-ray background, was modeled by fitting this component at high Galactic latitude (file ``iso\_source\_v06.txt''
provided with the {\em Fermi Science Tools}). The ``gll\_iem\_v06'' Galactic diffuse emission model \citep{2016ApJS..223...26A}
was used (the Galactic longitude and latitude of 3C~454.3 are 86$^{\circ}$.1 and −38$^{\circ}$.2, respectively).
All point sources in the third {\em Fermi}-LAT source catalog \citep[3FGL,][]{3FGL} located in the ROI and an additional surrounding 10$^{\circ}$ wide annulus (called ``source region'')
were modeled in the fits, with the spectral parameters kept free only for the three brightest point sources in the ROI (sources with a detection significance in 3FGL having a TS$>45$, namely 3C~454.3, CTA 102, and 3FGL J2243.9$+$2021) and for the isotropic and Galactic models. Depending on the type (spectral or time domain) of analysis, either the normalization only, or other spectral parameters of the three point sources, were kept free.

The source variability was investigated by producing light-curves with both 1 day and 3 hr binnings, and also by using an unbinned method for the study of fastest variability.
Although spectral breaks and curvatures were found to be characteristics of the actual spectral shape of 3C 454.3, light-curves were produced by modeling the spectra in each time bin as a simple PL over the 0.1--300 GeV energy range, since the statistics in these narrow time bins is not enough to significantly favor spectral shapes more complex than PL. Also, the statistical uncertainties of the photon indices obtained from PL fitted models are smaller than those obtained from these more complex functions \citep{Ackermann2010}. The PL fit function used in the {\em unbinned likelihood} procedure is defined in the following paragraph.

\begin{table*}[t!] 
\begin{center}
  \tabletypesize{\scriptsize}
  \scriptsize
  \caption{3C~454.3 soft X-ray spectral fit. \label{tab:3c454-fit}}
\begin{tabular}{cccccccccc}
  \tableline
  \tableline
  XRT Obs.      & Start    		& Expo &  $\Gamma$	& $\Gamma_{\rm hard}$      	& E$_{\rm break}$	& Norm ($10^{-3}$~ph	& $\chi^2-$ & dof & flux 2--6 keV\\
\scriptsize (id--modes) 		& \scriptsize(date) & \scriptsize(s) & 	& 			&
\scriptsize(keV)           	& $\times$~keV$^{-1}$~cm$^{-2}$~s$^{-1}$) 	& 	&	&  \scriptsize($10^{-11}$~erg~cm$^{-2}$~s$^{-1}$)\\
\tableline
00031018015 PC & 2014 Jun 14 06:23:59  & 3965 & 1.44$^{1.49}_{1.38}$ &  \nodata           &  \nodata           &  6.28$^{6.55}_{6.01}$  &  96.52 & 123 & 2.23$^{2.29}_{2.16}$ \\
               & (MJD 56822.27)        &                          &                    &                    &                      &              &\\
00031018016 PC & 2014 Jun 16 14:20:58  & 2988 & 1.54$^{1.60}_{1.48}$ &  \nodata           &  \nodata           &  6.78$^{7.12}_{6.44}$  &  90.56 & 94  & 2.10$^{2.17}_{2.04}$ \\
               &( MJD 56824.60)        &                          &                   &                    &                      &              &\\
00031018017 WT & 2014 Jun 17 16:10:59  & 1793 & 1.91$^{2.02}_{1.80}$ &  \nodata           &  \nodata           &  8.57$^{9.14}_{8.01}$  &  53.60 & 51  & 1.66$^{1.77}_{1.57}$ \\
               & (MJD 56825.67)        &                          &                   &                    &                      &              &\\
00031018019 WT & 2014 Jun 18 03:09:59  & 1584 & 2.33$^{2.53}_{2.15}$ &  1.53$^{1.68}_{1.38}$ &  1.62$^{1.88}_{1.40}$ &  9.18$^{9.66}_{8.69}$  &  65.51 & 67  & 1.97$^{2.01}_{1.89}$ \\
               & (MJD 56826.13)        &                          &                   &                    &                      &              &\\
00031018020 WT & 2014 Jun 21 14:19:59  & 1684 & 2.47$^{2.60}_{2.34}$ &  1.76$^{1.86}_{1.65}$ &  1.53$^{1.74}_{1.38}$ & 22.38$^{23.18}_{21.60}$ & 142.66 & 133 & 3.86$^{4.01}_{3.75}$ \\
               & (MJD 56829.60)        &                          &                   &                    &                      &              &\\
\tableline
\end{tabular}
\end{center}
\end{table*}

Spectral analysis has been processed over several epochs of the flaring activity and fits were performed over the 0.1--300 GeV range, using the unbinned likelihood analysis package. The four following functions were successively used in our analysis:

\begin{enumerate}
\item a {\em power law (PL)}, defined as 
  \begin{equation}
    dN(E)/dE=N_0~(E/E_{\rm p})^{-\Gamma}, \label{Eq:PL}
  \end{equation}
  with $E_p$ = 412.7 MeV, which is the value of the pivot energy given in 3FGL;
\item a {\em broken power law (BPL)}, defined as 
  \begin{equation}
    dN(E)/dE=N_0~(E/E_{\rm break})^{-\Gamma_i}, \label{Eq:BPL}
  \end{equation}
  with $i=1$ if $E<E_{\rm break}$ and $i=2$ if $E>E_{\rm break}$;
\item a {\em log-parabola (LP)}, defined as 
  \begin{equation}
    dN(E)/dE=N_0~(E/E_0)^{-\alpha-\beta~\ln(E/E_0)},\label{Eq:LP}
  \end{equation}
  with $E_0$ = 297.6 MeV, which is the value of the pivot energy of the {\em LP} spectral fit of 3C~454.3 as reported in the {\em Second} Fermi-{\em LAT source catalog (2FGL)} in \citet{2FGL}, and where ``$\ln$'' is the natural logarithm;
\item a {\em power law with an exponential cutoff (PLEC)}, defined as 
  \begin{equation}
    dN(E)/dE=N_0~(E/E_p)^{-\Gamma_{\rm PLEC}}~\exp(-E/E_{\rm c}), \label{Eq:PLEC}
  \end{equation}
  with $E_p$ = 412.7 MeV (as for the PL). 
\end{enumerate}

he estimated systematic uncertainty in the effective area is 5\% in the 100~MeV--100~GeV range. The energy resolution ($\Delta E/E$, at 68\% containment) is 20\% at 100 MeV, and between 6 and 10\% over the 1--500 GeV range.\footnote{http://fermi.gsfc.nasa.gov/ssc/data/analysis/LAT\_caveats.html} \footnote{http://www.slac.stanford.edu/exp/glast/groups/canda/lat\_\\Performance.htm}

\section{{\em Swift}-XRT analyses} \label{sec:Swift}

During the 2014 June outburst revealed by the LAT, five {\em Swift}-XRT
observations of 3C~454.3 were performed in two observation modes: photon
counting (PC) mode and windowed timing (WT) mode. Table~\ref{tab:3c454-fit} reports
the observation log.

\begin{figure*}[t!]
\epsscale{1.0}
\plotone{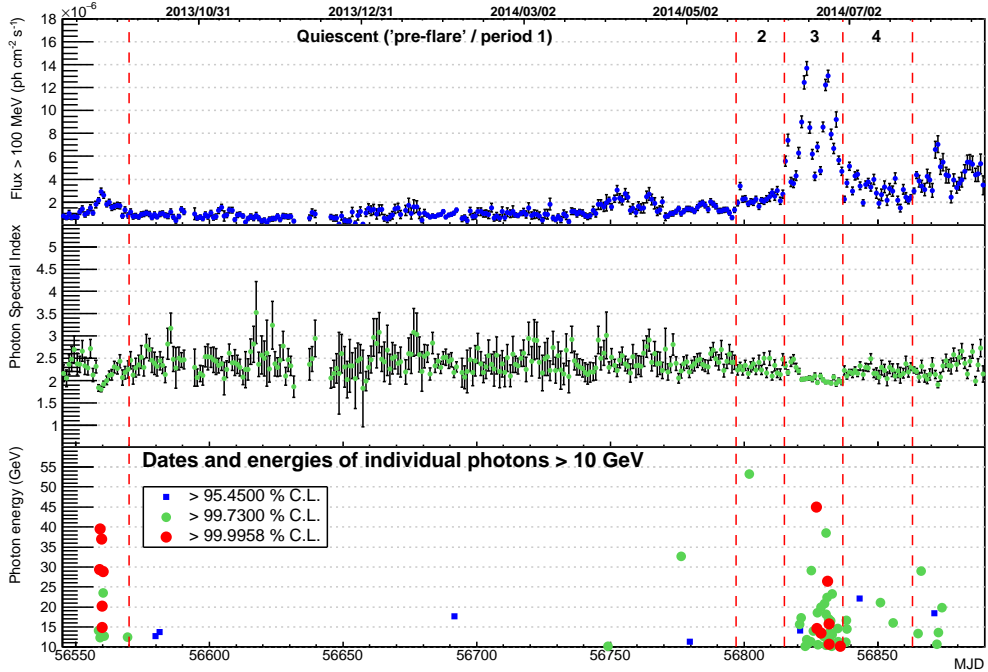}
\caption{Top panel: {\em Fermi}-LAT light-curve of 3C~454.3 from 2013 August to 2014 August, including the 2013 September 23--25 flare, the 2013 October 5--2014 May 20 pre-flare (phase 1) and the 2014 May--July outburst (phases 2, 3, 4), with 1 day binning. Gaps appeared in the light-curve during the pre-flare due to low-exposure modes. Middle panel: photon index ($\Gamma$) versus time. Bottom panel: arrival time and energy of E$>$10 GeV photons with three different significance levels of source association (2-, 3-, and 4-$\sigma$ Gaussian equivalent). Vertical dashed lines separate the four phases studied.\label{fig:4period_LC}}
\end{figure*}
\begin{table*}
\begin{center}
\caption{Outburst phases of 3C~454.3, as identified from the light-curves. \label{tbl-periods}}
\small
\begin{tabular}{llccr}
\tableline
\tableline
 & Phase & Dates & MJD & Duration (days)  \\
\tableline
1 & Pre-flare    & 2013 Oct 5--2014 May 20 & 56570.0--56797.0 & 227.0 \\
2  & Plateau     & 2014 May 20--Jun 7  & 56797.0--56815.0 &  18.0 \\
3  & Flare       & 2014 Jun 7--Jun 29 & 56815.0--56837.0 &  22.0 \\
4  & Post-flare  & 2014 Jun 29--Jul 25 & 56837.0--56863.0 &  26.0 \\
\tableline
3a & Flare I     & 2014 Jun 10--18      & 56818.5--56826.5 &   8.0 \\
3b & Flare II    & 2014 Jun 20--25      & 56828.0--56833.5 &   5.5 \\
\tableline
\end{tabular}
\end{center}
\end{table*}

Observations were analyzed using {\texttt{xrtproducts}} and {\em HEAsoft 6.15}. For the analysis,
events were extracted from a circular ROI centered on the source position having a radius
of $\sim$20 pixels \citep[that corresponds to $\sim$47 arcsec,][]{moretti04}. The background is
computed from a source-free nearby region having a radius of $\sim$50 pixels. Spectral data were
grouped to a minimum of 20 counts per energy bin that allows us to confidently use the chi-square
statistic.

In our spectral analysis approach, we first fit the spectra with the simplest model, which is a
PL keeping the absorption fixed to the Galactic value
($N_{H}^{gal}$ = 6.63 $\times$ 10$^{20}$ atoms cm$^{-2}$). This value is derived from the LAB
Survey of Galactic HI database \citep{kalberla05}. If the fit result can be improved with more
sophisticated models, we estimate its significance with the F-test \citep{bevington92}. As a
result we find that none of the spectra require absorption in addition to the Galactic value. This also
excludes variable absorption for 3C~454.3 because we have analyzed data from five
different epochs.
In more detail, observations 00031018014, 00031018016 and 00031018017 are best modeled by a PL. Instead a BPL
best reproduces the spectra of observations 00031018019 and 00031018020. For both observations the low energy spectral index is soft ($\Gamma$~$\sim$2.4), while above the break ($E_{\rm break}$~$\sim$1.5~keV) the spectral index hardens ($\Gamma_{\rm hard}$~$\sim$1.6).
A simple power-law model applied to these two observations leads to a fit result of $\chi^{2}$ =  88.98/69 degrees of freedom (dof)
for the former and $\chi^{2}$ =  183.14/135 dof for the latter observation. By applying the F-test
the result shows that the spectral fit improvement obtained by using the BPL model
with respect to the simple PL is not due to chance. While the fit improvement is significant, it is worth noting that these two spectra are of a lower quality since they have the lowest exposure ($\sim$1.6 ks) of the observations.
The spectra were analyzed with
XSPEC 12 \citep{arnaud96}. The fitting was performed over the energy band 0.6--6.0 keV.
Fluxes are computed in the 2.0--6.0 keV energy range. All errors are presented at 90\% CL.
The results are shown in Table~\ref{tab:3c454-fit}.


\section{Results}

In this section, we report on variability studies and spectral analysis of 3C~454.3 during the outburst phase of 2014 May--July (plateau, flare, post-flare), as well as during the ``pre-flare'' quiescent phase preceding the outburst, using \emph{Fermi}-LAT data.

\subsection{Light-curves} \label{subsec:LCs}

The characteristic temporal evolution of the flux of 3C~454.3 during its May--July outburst can be identified by a three-phase pattern (plateau, flare and post-flare), as described in the introduction. We show in Fig. \ref{fig:4period_LC} a 1-day light-curve encompassing the pre-flare and the three-phase outburst. These phases are described in Table \ref{tbl-periods}. A pre-flare phase was observed, from MJD 56570 to 56797, during which the flux remains, for most of the time, below $F_{100}=2$. This phase started a few days after the 2013 September 23--25 outburst, and lasted for 227 days. Analysis of the pre-flare data is used to show the spectral behavior and flux level during the quiescent phase of the source. A plateau phase was then observed from MJD 56797 to 56815, during which the flux rose slightly and remained quite stable. During the flare phase, from MJD 56815 to 56837, the flux rose dramatically and exhibited a complex, structured trend. A post-flare phase was observed, during which the flux was at a higher level than the plateau phase and remained fluctuating. We define the post-flare phase between MJD 56837 to 56863, before secondary outbursts show up again. Indeed, 3C~454.3 remained in a relatively high state for several weeks after this outburst episode. In this study, we also present a more detailed analysis of the flaring phase by scanning subflares on shorter timescales.

\begin{figure*}[t!]
\epsscale{1.0}
\plotone{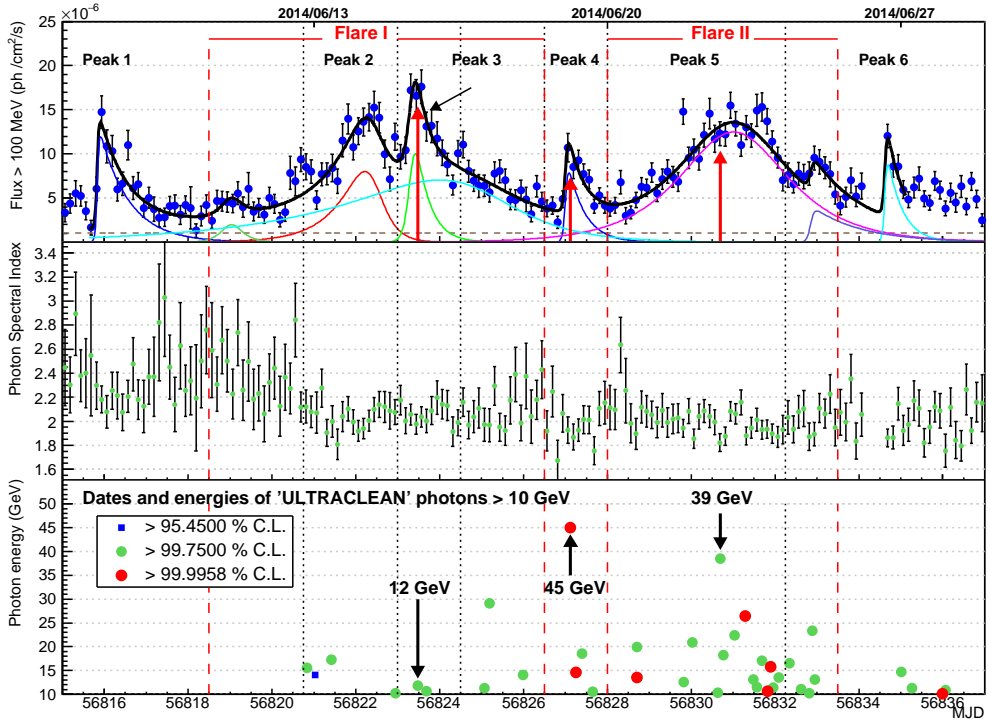}
\caption{Top panel: {\em Fermi}-LAT light-curve of the flare phase with 3 hr binning. Flaring peaks were fitted by the $F$ function (in Eq. \ref{eq:Eq_flare_fit}) for nine structures. Fits were performed in the MJD 56815.625--56835.130 time range. The thin color lines correspond to the contribution of single peaks in the total fit, which is represented by the thick black line. The dashed brown line is the fitted constant baseline, which also contributes to the total fit. Six major peaks were labeled for more detailed studies. The red arrows indicate the arrival time of the three high-energy photons used to calculate the Doppler factor in section \ref{sec:disc} (Discussion), and whose energies are labeled in the bottom panel. Due to an instrumental problem, the MJD 56834.375 bin contains no data. Middle panel: photon spectral index ($\Gamma$) of the PL fits of data. Bottom panel: arrival time and energy of $E>10$ GeV photons with three different significance levels of source association (2-, 3-, and 4-$\sigma$ Gaussian equivalent). This panel is a zoom of the bottom panel of Fig. \ref{fig:4period_LC}. Vertical red dash lines indicate the two major flaring phases (I and II), and black dotted lines indicate Peaks 2, 3, 4 and 5. \label{fig:flare_fits_Tr_Tf}}
\end{figure*}

Besides the flux evolution that defines an outburst, a spectral evolution is also observed, mainly characterized by a hardening of the photon spectral index (PL fitting), and detection of high-energy (HE) photons ($>$ 10 GeV). The photon indices and HE photons are displayed respectively in the middle and lower panels of Fig. \ref{fig:4period_LC}. A significant hardening of the photon spectral index is observed when the flux increases, with values of $\Gamma$ going below 2 during some parts of the flare. More precisely, we can report, from the 3 hr binned light-curve in Fig. \ref{fig:flare_fits_Tr_Tf}, that the photon index $\Gamma < 2$ for 75 \% of the time bins of Peak 4. We also report that $\Gamma < 2$ for $\sim$40 \% of the time bins of Peak 5, mainly occurring during the second half of the peak.

\begin{table*}
\begin{center}
\caption{Parameters of fit function $F$, for the 6 peaks identified in Fig. \ref{fig:flare_fits_Tr_Tf}. \label{tbl-LC_fits}}
\small
\begin{tabular}{lllll}
\tableline\tableline
Peak & $t_0$  & $F_0$                                     & $T_r$                   & $T_f$   \\
     & (MJD)  & \footnotesize $(10^{-6}$~ph~cm$^{-2}$~s$^{-1})$ & \footnotesize (hr)     & \footnotesize (hr)\\
\tableline
1   &    56815.8320  &    6.9  $\pm$   0.7  &   0.6  $\pm$  0.1  &   16.8 $\pm$   2.0\\
2   &    56822.3984  &    7.2  $\pm$   0.3  &   16.6 $\pm$  2.2  &   6.5  $\pm$   1.7\\
3   &    56823.3281  &    8.4  $\pm$   1.3  &   2.1  $\pm$  0.7  &   7.9  $\pm$   1.6\\
4   &    56827.0078  &    5.2  $\pm$   0.9  &   0.9  $\pm$  0.3  &   9.5  $\pm$   1.8\\
5   &    56831.0117  &    12.5 $\pm$   0.4  &   27.8 $\pm$  1.5  &   27.8  $\pm$  1.9\\
6   &    56834.6250  &    5.8  $\pm$   0.9  &   0.7  $\pm$  0.9  &   8.7  $\pm$   2.0\\
\tableline
\end{tabular}
\begin{tabular}{c}
  {\bf Note:} A constant baseline flux of $(1.05 \pm 0.05) \times 10^{-6}$~ph~cm$^{-2}~$s$^{-1}$\\
  was also fitted to the data. The $\chi^2/ndf$ for the whole fit is 183.4/115 = 1.60.
\end{tabular}
\end{center}
\end{table*}

In order to probe the flaring pattern over the whole flare phase (MJD 56815.0--56837.0) and to isolate subflaring events, we performed fits over different peaks observed in the 3 hr light-curve (Fig. \ref{fig:flare_fits_Tr_Tf}). 
The highest flux is reached during MJD 56823 (maximum of Peak 3), with the value $F_{100} = 17.6 \pm 1.9$, associated with PL index $\Gamma=2.0 \pm 0.1$, in a 3 hr bin, at 56823.5625. 
Note that the hard spectra with $\Gamma<2$ are observed at the end of Peak 4, 5 and 6. The lowest value of the photon index (also measured in a 3 hr bin) was $\Gamma = 1.7 \pm 0.2$, at MJD 56826.8125. Similar values ($\lesssim 1.8$) were also obtained at other time bins\footnote{at MJD 56821.5625, 56827.6875, 56830.6875, 56833.9375, 56835.5625, 56836.0625, 56836.3125 and 56836.4375} (see Fig. \ref{fig:flare_fits_Tr_Tf}).

\begin{figure*}[t!]
\epsscale{1.0}
\plotone{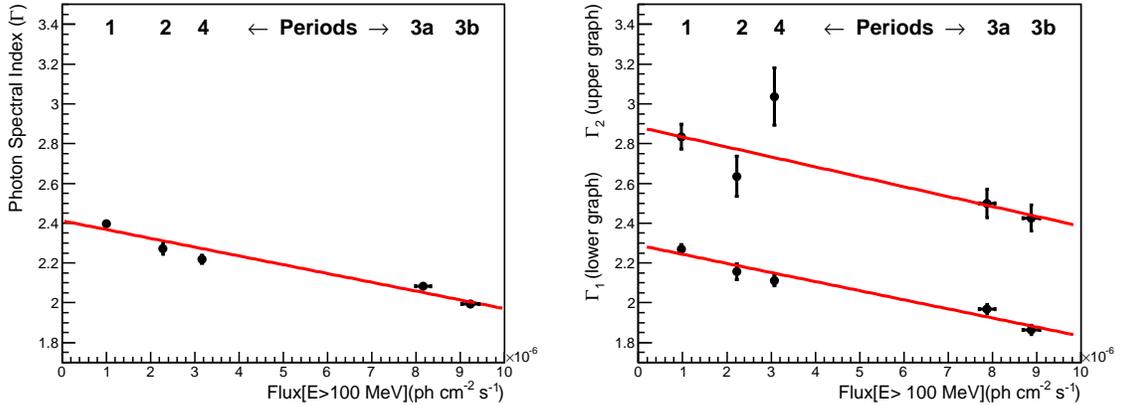}
\caption{Photon spectral index versus $F_{100}$ for five phases of the outburst: pre-flare, plateau, flare I, flare II and post-flare, as defined in Table \ref{tbl-periods}. Left panel: fluxes are calculated with PL model of photon index $\Gamma$. Right panel: fluxes are calculated with BPL model of photon indices $\Gamma_1$ and $\Gamma_2$. Values are reported in Table \ref{tab:fits}.} \label{fig:Gamma_vs_Flux_periods}
\end{figure*}

Through a detailed observation of the 3 hr light-curve we can identify nine individual flaring structures with a flux above $5 \times 10^{-6}$ ph~cm$^{-2} $s$^{-1}$. We fitted these nine peaks in order to account for the whole light-curve pattern over the MJD 56814.5--56835.1 range, in addition to a constant function to represent the baseline flux. Each of the nine fits was performed using the following function:
\begin{equation} \label{eq:Eq_flare_fit}
F=2 F_0(e^{(t_0-t)/T_r} + e^{(t-t_0)/T_f})^{-1},
\end{equation}

where:

\begin{enumerate}
\item $t_0$ is the time of the peak value;
\item $T_r$ and $T_f$ are the rise and fall time, respectively;
\item $F_0$ is the flux at $t_0$, representing the subflare amplitude.
\end{enumerate}

However, in the following discussions, we will only consider the six major peaks indicated in Fig. \ref{fig:flare_fits_Tr_Tf}. From our best fit model, we obtained a rise time $T_r$= 0.9$\pm$0.3 hr, associated with Peak 4, centered at $t_0=56827.0078$ (Table \ref{tbl-LC_fits}).
Peaks 2 and 5 are the broadest peaks, both reaching $F_{100} \simeq 14 \pm 2$ with their $T_r$ and $T_f$ of the order of one day. Peak 1 and Peak 4 are the objects of a more detailed study in Section \ref{subsec:FatestVar}, when we investigate the fastest variability pattern.

We also observe in Fig. \ref{fig:flare_fits_Tr_Tf} more details about the variation of the photon index during the flare. The ``Flare~I'' and ``Flare~II'' labels identify the two major subflares that will be studied in more detail later in this section, and that we have already defined in Table \ref{tbl-periods}. Time ranges of Peaks 2, 4 and 5 are indicated by dashed/dotted vertical lines.

\begin{table*}
\begin{center}
  \caption{Parameters of the PL, BPL, LP and PLEC fit functions obtained from the spectral analysis using the likelihood analysis method, for Peaks 2, 3, 4 and 5.} \label{tab:fits_peaks}
  \scriptsize
\begin{tabular}{cccccccc}
  \tableline
  \tableline
  \em \bf{PL}   & Date &   $F[>100$ MeV]     &         &  $\Gamma$     &            &                       & $-$Log(likelihood)          \\
                & (MJD) & $(10^{-6}$~ph~cm$^{-2}$~s$^{-1})$  &  &                   &                     &                       &           \\             
  \tableline
  Peak 2 & 56820.75--56823.00  & 10.5 $\pm$ 0.4   & &  2.04 $\pm$   0.03 &	       &			  &    13972.7\\
  Peak 3 & 56823.00--56824.50  & 12.0 $\pm$ 0.5   & &  2.05 $\pm$   0.03 &	       &			  &	9957.5\\
  Peak 4 & 56826.50--56828.00  & 5.9  $\pm$ 0.3   & &  1.97 $\pm$   0.04 &	       &			  &	8165.5\\
  Peak 5 & 56828.00--56832.25  & 9.6  $\pm$ 0.2   & &  1.98 $\pm$   0.02 &	       &			  &    30638.6\\
  \tableline
  \em \bf{BPL} & Date &    $F[>100$ MeV]    & Luminosity    &   $\Gamma_1$    &  $\Gamma_2$	  & $E_{\rm break}$      &	 $\Delta$Log(likelihood)\\
  & (MJD) & ($10^{-6}$~ph~cm$^{-2}$~s$^{-1}$)  &  ($10^{48}$~erg~s$^{-1}$) &                  &             &                       &        \\       
  \tableline
  Peak 2 & 56820.75--56823.00  & 10.2 $\pm$ 0.4   & 30.9 &  1.93 $\pm$   0.04 &  2.47 $\pm$   0.12   &   $2000_{-400}^{+1000}$   &  $-$9.6 \\
  Peak 3 & 56823.00--56824.50  & 11.4 $\pm$ 0.5   & 33.7 &  1.86 $\pm$   0.05 &  2.44 $\pm$   0.10   &    $1100_{-200}^{+400}$   & $-$11.7  \\
  Peak 4 & 56826.50--56828.00  & 5.6  $\pm$ 0.3   & 19.7 &  1.82 $\pm$   0.06 &  2.43 $\pm$   0.16   &    $1900_{-400}^{+800}$   & $-$6.7 \\
  Peak 5 & 56828.00--56832.25  & 9.2  $\pm$ 0.2   & 30.8 &  1.85 $\pm$   0.03 &  2.40 $\pm$   0.07   &    $1700_{-200}^{+300}$   & $-$27.2 \\
  \tableline
  \em \bf{LP}   & Date &     $F[>100$ MeV]       & Luminosity      &   $\alpha$        & $\beta$	  &         & $\Delta$Log(likelihood)\\
  & (MJD) & $(10^{-6}$~ph~cm$^{-2}$~s$^{-1})$  & ($10^{48}~erg~s^{-1}$) &                   &                   &                      & \\             
  \tableline
  Peak 2 & 56820.75--56823.00  & 10.1 $\pm$ 0.4   & 31.1  &  1.88 $\pm$   0.05 &  0.08 $\pm$   0.02   & 			 & $-$9.8  \\
  Peak 3 & 56823.00--56824.50  & 11.3 $\pm$ 0.5   & 33.7 &  1.85 $\pm$   0.06 &  0.10 $\pm$   0.02   & 			 &  $-$11.3   \\
  Peak 4 & 56826.50--56828.00  & 5.6  $\pm$ 0.3   & 20.3 &  1.78 $\pm$   0.07 &  0.08 $\pm$   0.03   & 			 &  $-$5.5   \\
  Peak 5 & 56828.00--56832.25  & 9.1  $\pm$ 0.2   & 30.5 &  1.79 $\pm$   0.03 &  0.09 $\pm$   0.01   & 			 &  $-$31.2 \\
  \tableline
  \em \bf{PLEC}  & Date       &  $F[>100$ MeV]     & Luminosity     &   $\Gamma_{\rm PLEC}$     & $E_{\rm c}$	  &  & $\Delta$Log(likelihood)\\
  & (MJD) & $(10^{-6}$~ph~cm$^{-2}$~s$^{-1})$  & ($10^{48}~erg~s^{-1}$)  &           &      &        &            \\             
  \tableline
  Peak 2 & 56820.75--56823.00  & 10.2 $\pm$ 0.4   & 30.9 &  1.92 $\pm$   0.04 &  14000 $\pm$   4600    & 			 &  $-$9.8  \\
  Peak 3 & 56823.00--56824.50  & 11.5 $\pm$ 0.5   & 32.7 &  1.86 $\pm$   0.05 &   7900 $\pm$   2300    & 			 & $-$12.5   \\
  Peak 4 & 56826.50--56828.00  & 5.7  $\pm$ 0.3   & 20.3 &  1.84 $\pm$   0.05 &  15900 $\pm$   7100    & 			 &  $-$5.4  \\
  Peak 5 & 56828.00--56832.25  & 9.2  $\pm$ 0.2   & 30.7 &  1.85 $\pm$   0.03 &  13300 $\pm$   2600    & 			 &  $-$27.8  \\
  \tableline
  \multicolumn{8}{l}{{\bf Note.} These parameters are defined in Equations (1)--(4). The quality of unbinned fits is given by the Log(likelihood) for each of these three }\\
  \multicolumn{8}{l}{fitting functions, for the PL model, along with other fit parameters. For the LP, BPL and PLEC models, values of $\Delta Log(likelihood)$ are given}\\
    \multicolumn{8}{l}{with respect to Log(likelihood) of the PL fit. Luminosity was estimated according to the formula of Equation (\ref{eq:L}).}
\end{tabular}
\end{center}
\end{table*}

For the four spectral models defined in Section \ref{sec:obs_ana}, the total integrated flux ($F_{100}$) has been computed for each phase of the outburst (1, 2, 3, 3a, 3b, 4) and for Peaks 2, 4 and 5 already mentioned. We show in Fig. \ref{fig:Gamma_vs_Flux_periods} the plots of the dependence of the photon spectral index $\Gamma$ on the flux for phases 1, 2, 3a, 3b and 4, for the PL fit (left panel) and BPL fit (right panel). The ``harder-when-brighter'' spectral property we already mentioned is clearly observable through these different phases of the outburst. It is reflected on the $\alpha$ and all $\Gamma$ indices of the four fitting functions (see Table \ref{tab:fits}). In particular, both $\Gamma_1$ and $\Gamma_2$ of the BPL model were lowered by the same amount $\Delta \Gamma \sim$0.3 (from pre-flare to flare), as well as the $\Gamma$ and $\Gamma_{\rm PLEC}$ indices of PL and PLEC functions respectively.

We searched for some time-dependent or geometrical pattern of the ``photon index versus flux[$>E_0$]'' graphs for Peaks 2, 3 and 4, in both 6 hr and 12 hr binning for each peak respectively, but we could not find any obvious one, the ``harder-when-brighter'' tendency being only partially observed in such narrow time intervals.

\subsection{HE photons and peak luminosity} \label{subsec:HE_L}

A harder spectral index associated with the detection of HE photons during the flare and post-flare phases seems to be a common feature of high-state activity for 3C~454.3.

In the bottom panel of both Fig. \ref{fig:4period_LC} and \ref{fig:flare_fits_Tr_Tf} are presented the HE photons above 10 GeV along with the light-curves. To identify $\gamma$-ray induced events with a high degree of accuracy, we use the ``ULTRACLEAN" class of events. We report events at energies $E>10~$GeV, along with their arrival time and a probability $>$0.9545 (2-$\sigma$ Gaussian equivalent) to be associated with 3C~454.3, within a ROI = 0$^\circ$.5. On Fig. \ref{fig:4period_LC} HE photons are shown for a one-year period, where the HE photons detected during the 2013 September 23--25 outburst are also displayed. We note that a 53 GeV photon was detected during the plateau phase, at MJD 56801.94, about 3.5 days after a small peak in the light-curve. The way HE photons are distributed during the flare phase (phase 3) is clearly visible in Fig. \ref{fig:flare_fits_Tr_Tf}. Most of the HE photons are detected during the second part of this phase, mainly after Peak 3. No HE photons were detected during Peak 1. Moreover, the bulk of HE photons is detected in the second half of the broad structures, as in both the Peak 2--Peak 3 and Peak 5 subflares. Around the maximum of Peak 5, we found one photon at E = 39 GeV (MJD 56830.6955) with a probability 0.999951, and around the maximum of Peak 4, one photon at E = 45 GeV (MJD 56827.1233) with a probability 0.999952.

In section \ref{subsec:LCs} we presented studies of the flaring pattern of Peaks 2, 3, 4, and 5, as shown in Fig. \ref{fig:flare_fits_Tr_Tf}. Results of the spectral analysis are presented in Table \ref{tab:fits_peaks}, along with estimation of the $\gamma$-ray luminosity corresponding to these integrated subflare phases.

The luminosity of the source was calculated for each of the three spectral shapes (BPL, LP and PLEC). Since we found photons up to the 40-60 GeV energy bin in the analysis of six years of 3C~454.3's data \citep{britto}, we chose to calculate the luminosity $L$ in the 0.1--60 GeV energy range. This integrated apparent isotropic luminosity $L$ was calculated according to the following formula:

\begin{equation}
L = 4 \pi d_{L}^{2} \int_{E_1}^{E_2} E \frac{dN(E)}{dE} dE, \label{eq:L}
\end{equation}
with $E_1=100$ MeV, $E_2=60$ GeV, the luminosity distance $d_L = 5.55$ Gpc = $1.71 \times 10^{28}$~cm, using the cosmological parameters defined in Section 1, and $dN(E)/dE$ the differential form of the spectral model we have used. The highest values of the luminosity are found for Peak 3, at $L \simeq 33 \times 10^{48}$~erg~s$^{-1}$, even peaking at $\sim 60 \times 10^{48}$~erg~s$^{-1}$ in the 3 hr time bin centered at MJD 56823.5625.

\subsection{Fastest variability} \label{subsec:FatestVar}

Some flares visible in the light-curves of Fig. \ref{fig:flare_fits_Tr_Tf} exhibit characteristic times commensurate with or shorter than the orbiting period of the \emph{Fermi} satellite (1.5 hr). The standard analysis whereby the photons are binned in
time is inappropriate for a detailed temporal characterization of these flares.
 In survey mode, a typical source is seen by the LAT only 20\%
of the survey period, $T_{surv}$, which is twice the orbiting period. The shortest time-binning commonly used in light-curves is $T_{surv}$ to avoid
explicitly dealing with LAT's discontinuous exposure pattern, but the reliability of the so-obtained
timescales is questionable when they are significantly shorter than the bin size. Instead of the standard analysis, we have used a maximum-likelihood approach \citep[described in the context of the adaptive-binning method, ABM, in][]{Lott} where a time-dependent function is fitted to the unbinned data (i.e., neither binned in time nor in energy).
\begin{figure*}[t!]
\epsscale{1.}
\plottwo{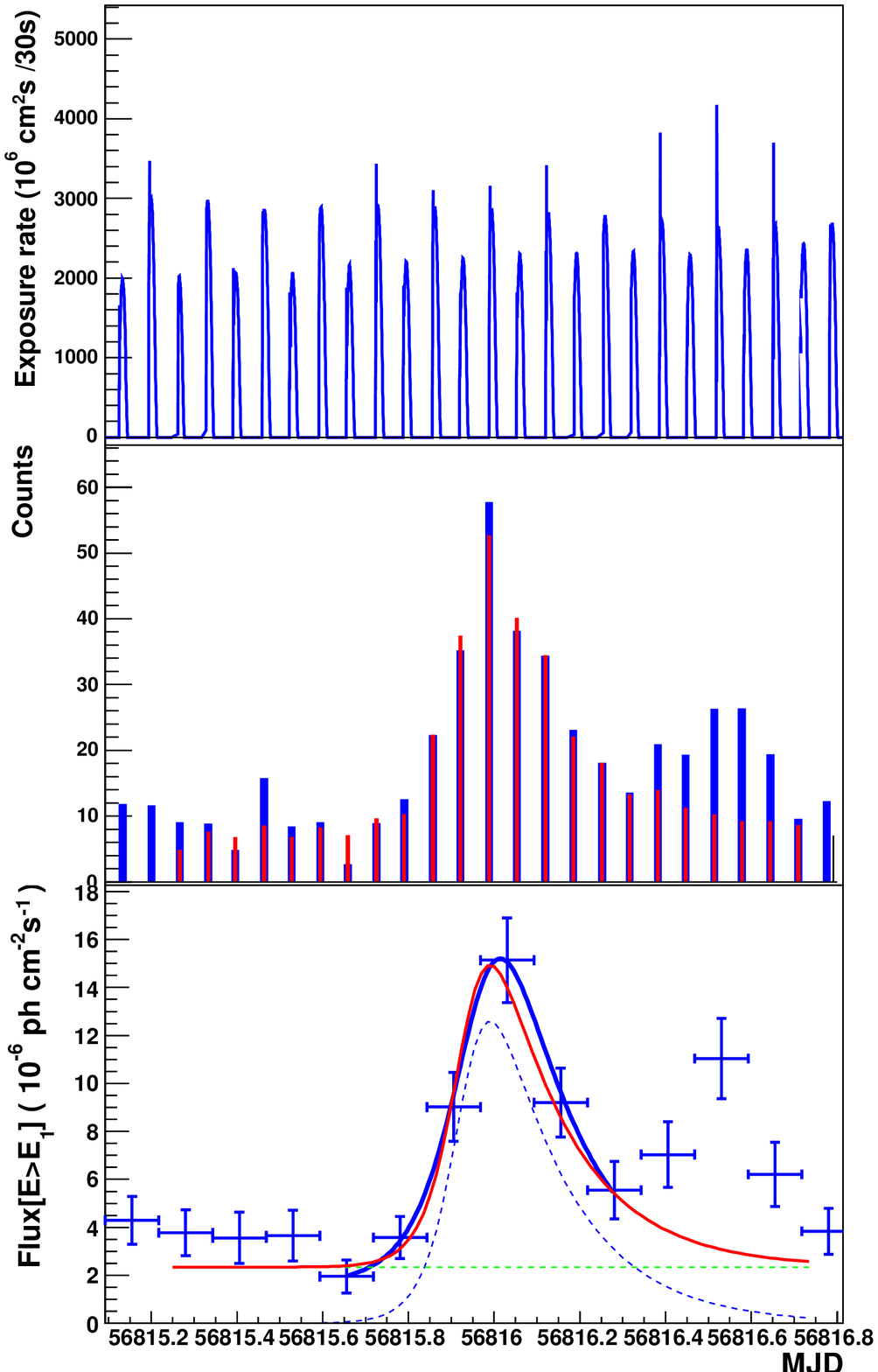}{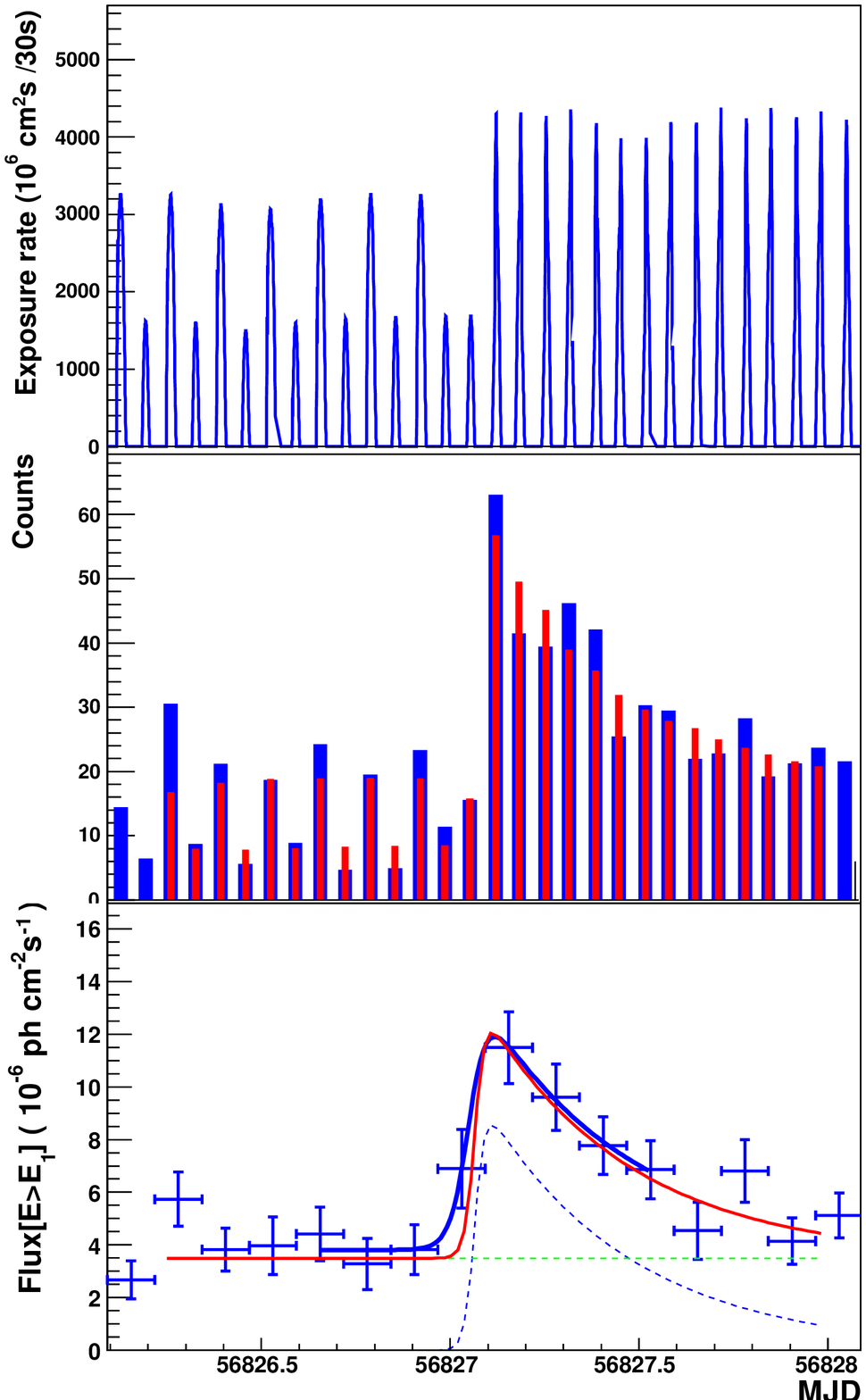}
\caption{Left: MJD 56816 (``Peak 1'') flare. Top panel: exposure rate as a function of time. Middle panel: measured (probability-weighted, blue) and modeled (red) photon counts. Bottom panel: light-curves. Blue points: 3 hr binning. Blue curve: function fitted to 3 hr points. Red curve: function fitted using the unbinned method. The green dashed line corresponds to the constant component and the blue one to the flare component $F(t)$ given in Eq. \ref{eq:Eq_flare_fit}. Right: same for the MJD 56827 (``Peak 4'') flare.} \label{fig:fast}
\end{figure*}
We used a function of the form $S_S(E,t)=S(E) \times (F(t) + B)$,
where $F(t)$ is given in Eq. \ref{eq:Eq_flare_fit}, $B$ is a constant component and $S(E)$ is a PL distribution (with fixed photon index, hence neglecting the spectral changes during the flare). To compute the likelihood, the needed instantaneous exposure rate was interpolated from the values assessed every 30 s (same time steps as in the spacecraft data file provided by the {\em Fermi Science Support Center}). In this section, we focus on the Peak-1 and -4 flares as they can be better characterized than the other fast flares.

\begin{table*}
\begin{center}
\scriptsize
\caption{Parameters of the PL, BPL, LP and PLEC fit functions obtained from the spectral analysis by the likelihood analysis, for the pre-flare and the three phases of the 2014 May--July outburst.} \label{tab:fits}
\begin{tabular}{cccccccc}
\tableline
\tableline
  \em \bf{PL}   & Date       &      $F[>100$ MeV]                &              &  $\Gamma$ &     & 	 &    $-$Log(likelihood)\\
                & (MJD)       & $(10^{-6}$~ph~cm$^{-2}$~s$^{-1})$  &   &           &      &          & \\             
\tableline
Pre-flare (1)  &  56570.0--56797.0  & 1.0 $\pm$ 0.02   & &  2.40 $\pm$   0.01 &		    &				&   406738.8\\
Plateau   (2)  &  56797.0--56815.0  & 2.3 $\pm$ 0.1    & &  2.27 $\pm$   0.03 &		    &				&    47820.6\\
Flare     (3)  &  56815.0--56837.0  & 7.5 $\pm$ 0.1    & &  2.04 $\pm$   0.01 &		    &				&   115792.8\\
Post-flare (4) &  56837.0--56863.0  & 3.2 $\pm$ 0.1    & &  2.22 $\pm$   0.02 &		    &				&    61635.6\\
\tableline 
Flare I  (3a)  & 56818.5--56826.5   & 8.2 $\pm$ 0.2    & &  2.08 $\pm$   0.02 &		    &				&    42023.7\\
Flare II (3b)  & 56828.0--56833.5   & 9.2 $\pm$ 0.2    & &  1.99 $\pm$   0.02 &		    &				&    37178.5\\
\tableline
\tableline
   \em \bf{BPL}  & Date         &   $F[>100$ MeV]  & Luminosity            &   $\Gamma_1$        &  $\Gamma_2$	      & $E_{\rm break}$		    & $\Delta$Log(likelihood)\\
               & (MJD)  & $(10^{-6}$~ph~cm$^{-2}$~s$^{-1})$  & ($10^{48}$~erg~s$^{-1}$)  &                   &                     &                              &             \\             
\tableline
Pre-flare (1)  & 56570.0--56797.0  & 1.0 $\pm$ 0.02   & 1.7  &  2.27 $\pm$   0.02 &  2.83 $\pm$  0.06	 &   $1100_{-100}^{+200}$ &  $-$34.9 \\
Plateau   (2)  & 56797.0--56815.0  & 2.2 $\pm$ 0.1    & 4.7  &  2.16 $\pm$   0.04 &  2.64 $\pm$  0.10	 &   $1100_{-300}^{+500}$ &   $-$8.9 \\
Flare     (3)  & 56815.0--56837.0  & 7.3 $\pm$ 0.1    & 21.8 &  1.93 $\pm$   0.01 &  2.50 $\pm$  0.04	 &   $2000_{-200}^{+200}$ &  $-$72.5 \\
Post-flare (4) & 56837.0--56863.0  & 3.1 $\pm$ 0.1    & 6.6  &  2.11 $\pm$   0.02 &  3.04 $\pm$  0.14	 &   $2100_{-500}^{+300}$ &  $-$26.6 \\
\tableline
Flare I  (3a)  & 56818.5--56826.5  & 7.9 $\pm$ 0.2    & 22.0 &  1.97 $\pm$   0.02 &  2.40 $\pm$  0.07	 &   $1700_{-300}^{+300}$ &  $-$23.7 \\
Flare II (3b)  & 56828.0--56833.5  & 8.9 $\pm$ 0.2    & 29.3 &  1.86 $\pm$   0.02 &  2.43 $\pm$  0.06	 &   $1800_{-200}^{+300}$ &  $-$32.4 \\
\tableline
\tableline
  \em \bf{LP}   & Date        &     $F[>100$ MeV] & Luminosity                   &   $\alpha$     & $\beta$    &  & $\Delta$Log(likelihood) \\
                & (MJD)       & $(10^{-6}$~ph~cm$^{-2}$~s$^{-1})$  & ($10^{48}$~erg~s$^{-1}$)  &              &            &  &            \\             
\tableline
Pre-flare (1)  & 56570.0--56797.0  & 1.0 $\pm$ 0.02   & 1.7  &  2.26 $\pm$   0.02 &  0.11 $\pm$  0.01	 &		      &   $-$35.7\\
Plateau   (2)  & 56797.0--56815.0  & 2.2 $\pm$ 0.1    & 4.7  &  2.16 $\pm$   0.04 &  0.08 $\pm$  0.02	 &		      &   $-$8.0\\
Flare     (3)  & 56815.0--56837.0  & 7.2 $\pm$ 0.1    & 21.9 &  1.88 $\pm$   0.02 &  0.08 $\pm$  0.01	 &		      &  $-$75.9\\
Post-flare (4) & 56837.0--56863.0  & 3.0 $\pm$ 0.1    & 6.7  &  2.05 $\pm$   0.03 &  0.11 $\pm$  0.02	 &		      &   $-$26.7\\
\tableline
Flare I  (3a)  & 56818.5--56826.5  & 7.8 $\pm$ 0.2    & 21.9 &  1.92 $\pm$   0.03 &  0.09 $\pm$  0.01	 &		      &  $-$27.7 \\
Flare II (3b)  & 56828.0--56833.5  & 8.8 $\pm$ 0.2    & 29.4 &  1.81 $\pm$   0.03 &  0.09 $\pm$  0.01	 &		      &  $-$34.4 \\
\tableline
\tableline
\em \bf{PLEC} & Date          &    $F[>100$ MeV] & Luminosity            &   $\Gamma_{\rm PLEC}$          & $E_{\rm c}$	      &		   & $\Delta$Log(likelihood)\\
              & (MJD)         & ($10^{-6}$~ph~cm$^{-2}$~s$^{-1})$  & ($10^{48}$~erg~s$^{-1}$)  &           &      &        &           \\             
\tableline
Pre-flare (1)  & 56570.0--56797.0  & 1.0 $\pm$ 0.02   & 1.7  &  2.25 $\pm$   0.03 &   6100 $\pm$  1100     &		       &   $-$31.4 \\
Plateau   (2)  & 56797.0--56815.0  & 2.2 $\pm$ 0.1    & 4.6  &  2.16 $\pm$   0.04 &  10300 $\pm$  3800     &		       &   $-$7.3  \\
Flare     (3)  & 56815.0--56837.0  & 7.3 $\pm$ 0.1    & 21.7 &  1.91 $\pm$   0.02 &  12700 $\pm$  1500     &		       &   $-$75.6 \\
Post-flare (4) & 56837.0--56863.0  & 3.1 $\pm$ 0.1    & 6.6  &  2.05 $\pm$   0.04 &   6300 $\pm$  1300     &		       &   $-$24.2 \\
\tableline	      
Flare I  (3a)  & 56818.5--56826.5  & 7.9 $\pm$ 0.2    & 21.8 &  1.95 $\pm$   0.03 &  11800 $\pm$  2400     &		       &   $-$25.5 \\
Flare II (3b)  & 56828.0--56833.5  & 8.9 $\pm$ 0.2    & 29.4 &  1.86 $\pm$   0.02 &  13400 $\pm$  2400     &		       &   $-$32.4 \\
\tableline
\multicolumn{8}{l}{{\bf Note.} These parameters are defined in Eq. 1-4. The quality of unbinned fits is given by the Log(likelihood) for each of these three}\\
\multicolumn{8}{l}{fitting functions, for the PL model, along with other fit parameters. For the LP, BPL and PLEC models, values of $\Delta$Log(likelihood) are given,}\\
\multicolumn{8}{l}{with respect to Log(likelihood) of the PL fit. Luminosity was estimated according to the formula of Equation (\ref{eq:L}).}
\end{tabular}
\end{center}
\end{table*}
%
\begin{figure*}[t!]
\epsscale{1}
\plottwo{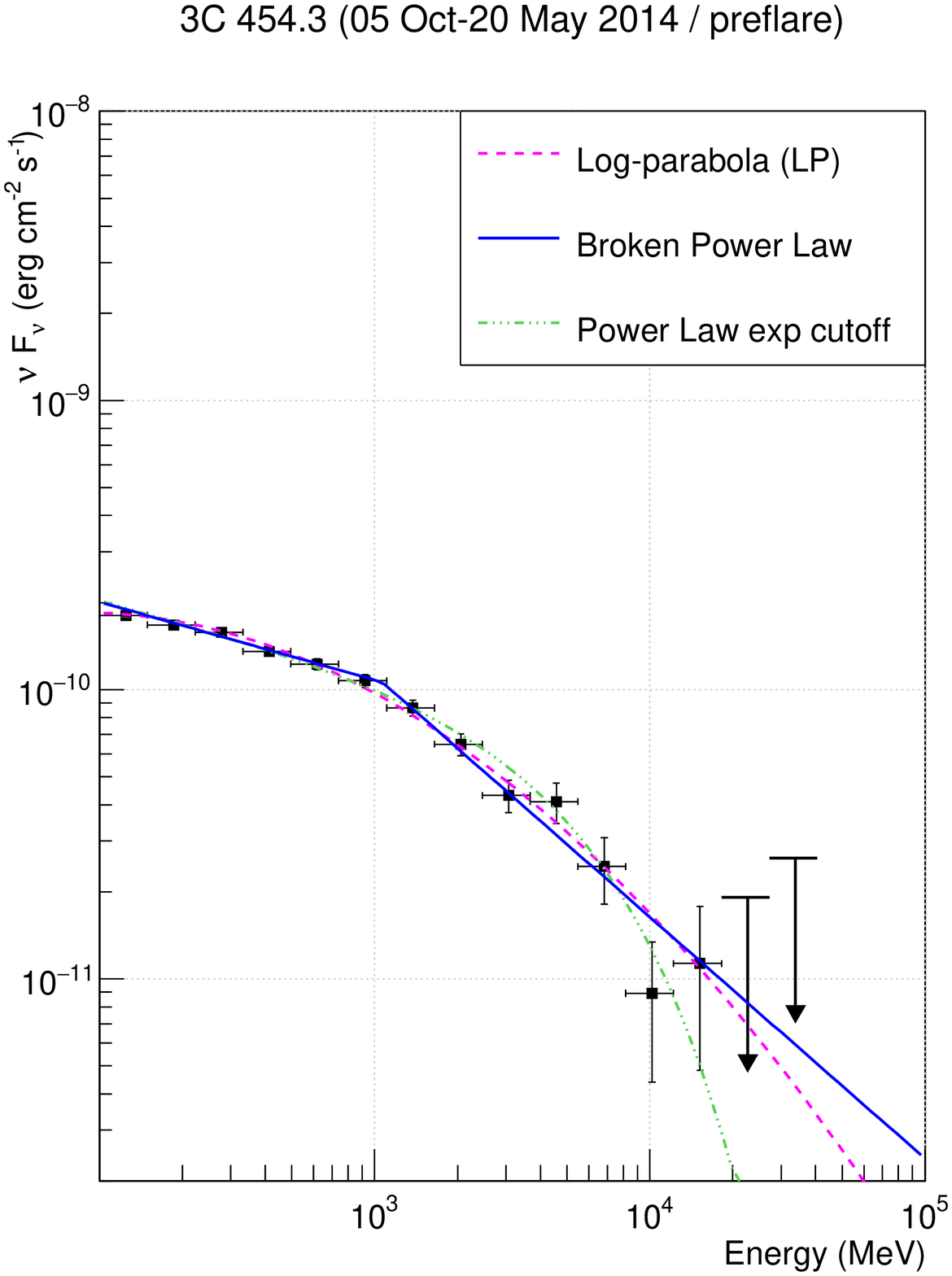}{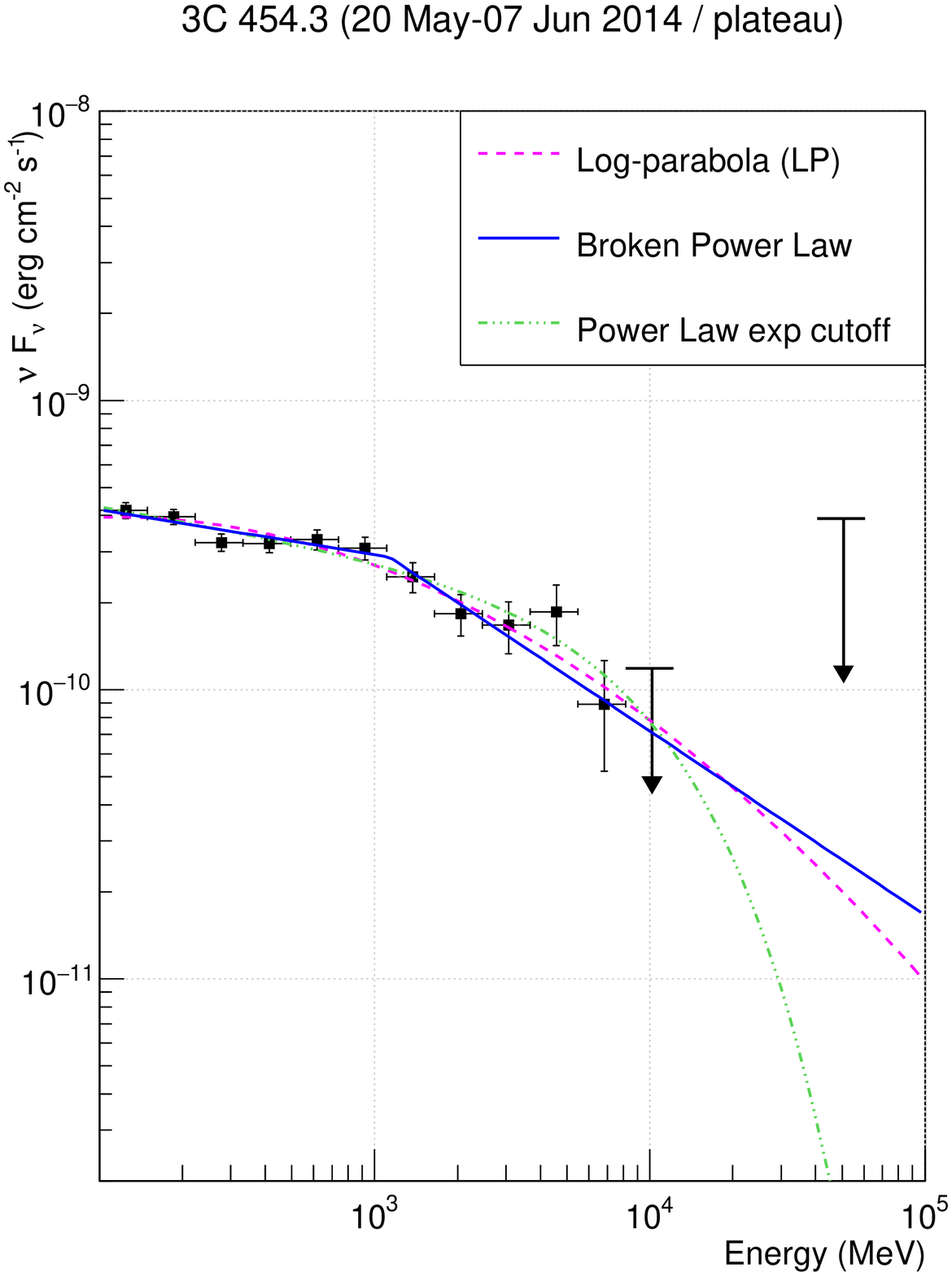}\\
\plottwo{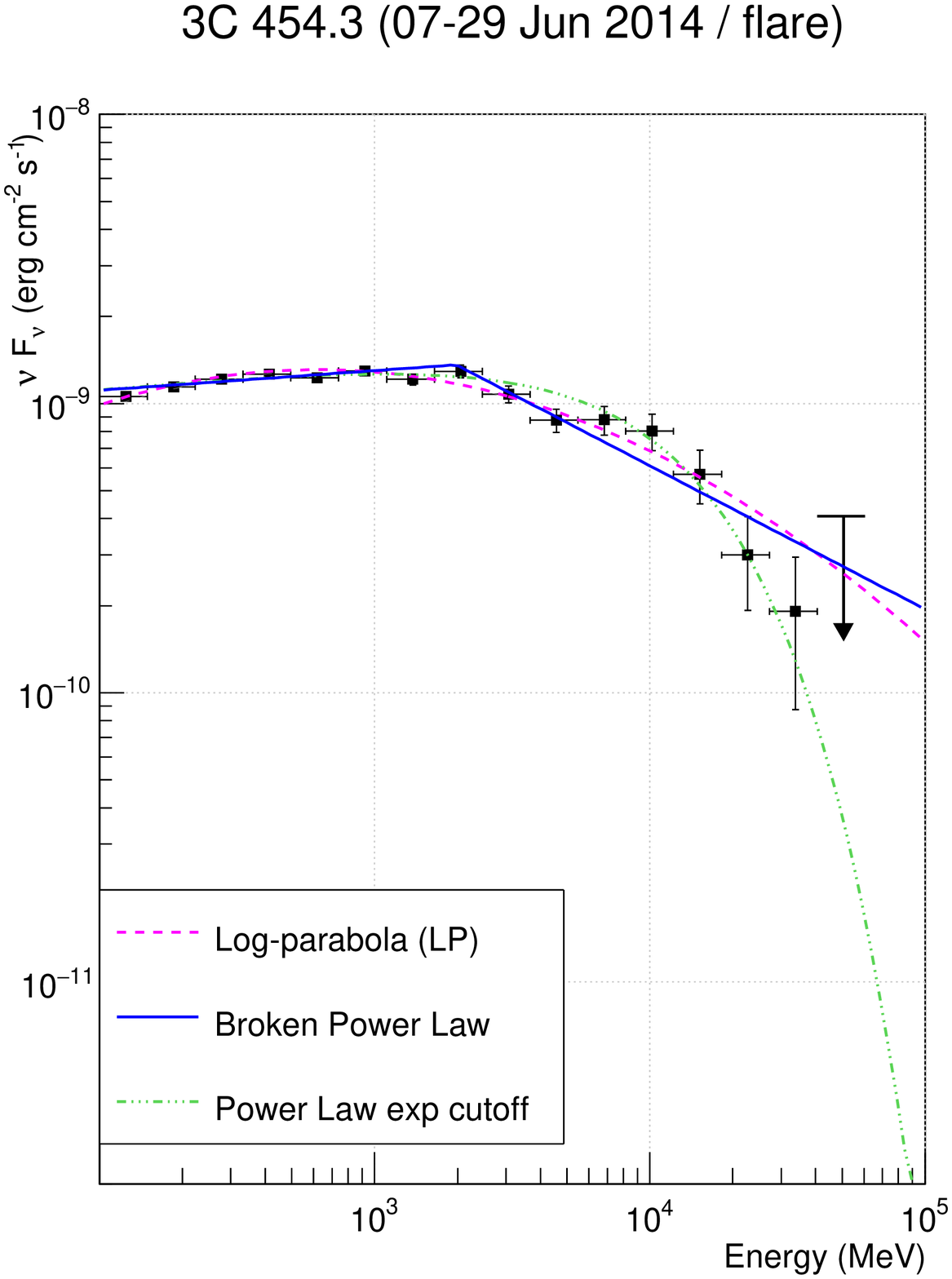}{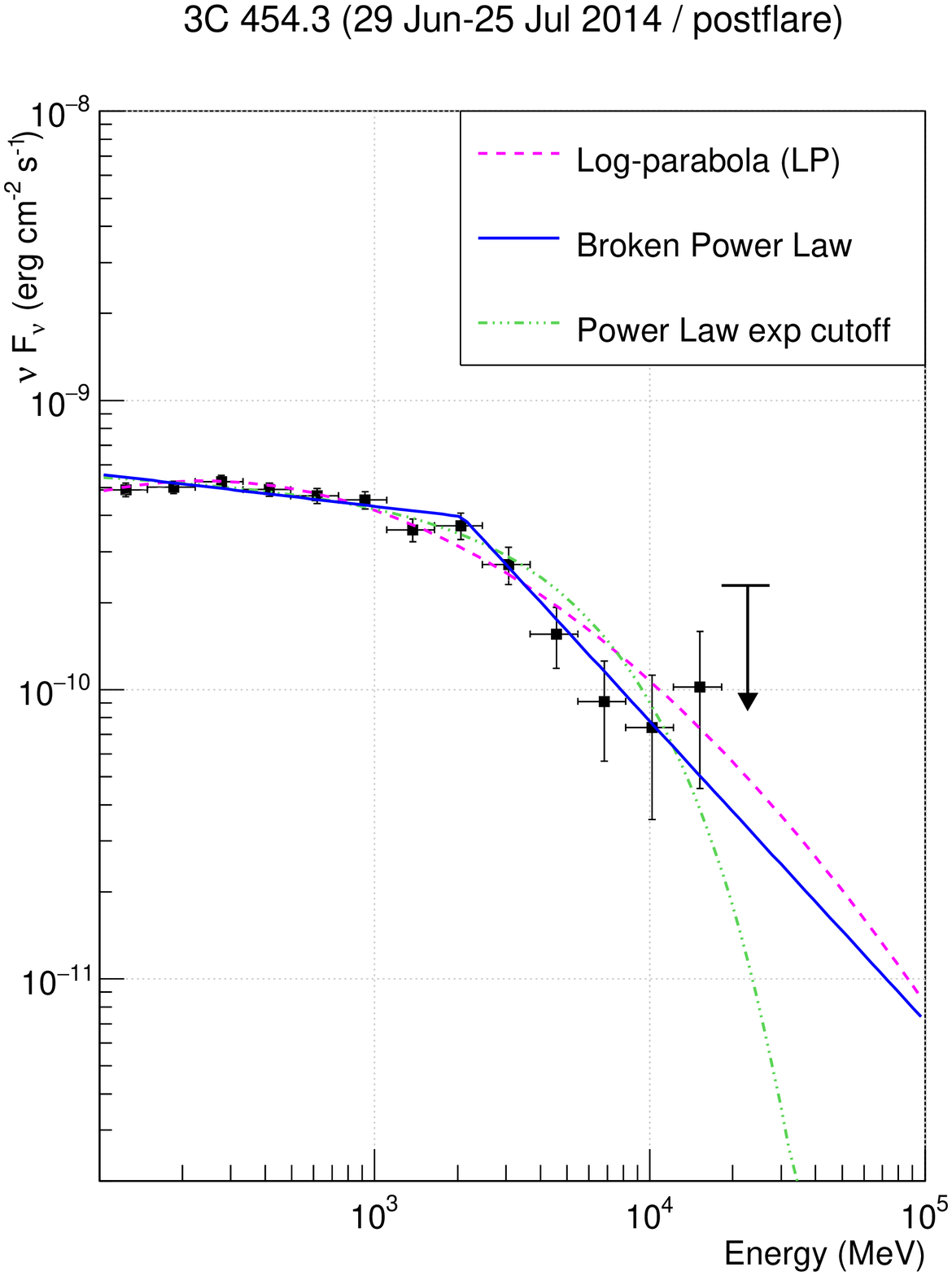}
\caption{Spectral energy distributions of 3C~454.3 above 100 MeV for the pre-flare, plateau, flare and post-flare phases, along with the LP, BPL and PLEC fitted functions. \label{fig:SEDs}}
\end{figure*}
\begin{figure*}[t!]
\epsscale{1}
\plottwo{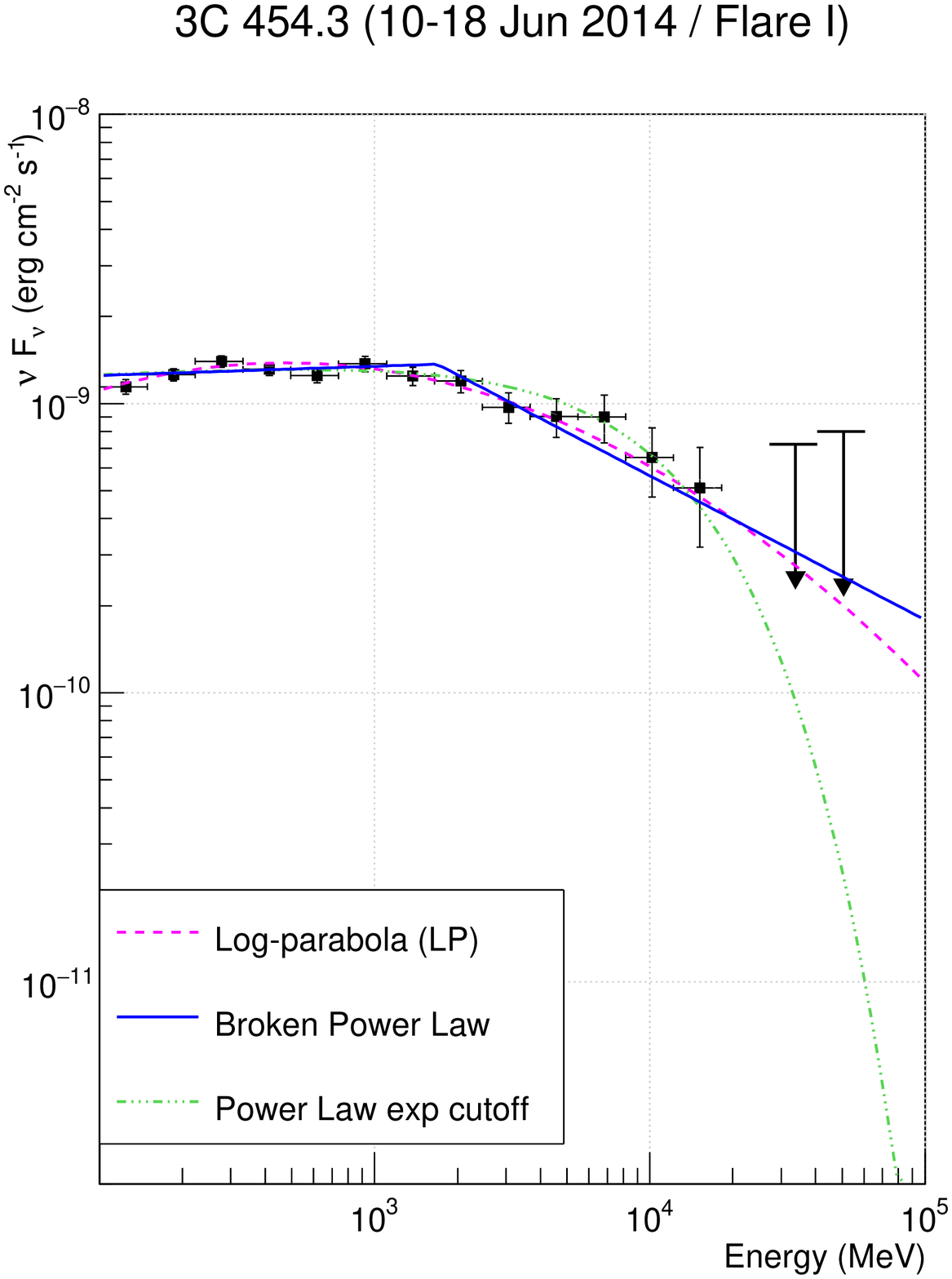}{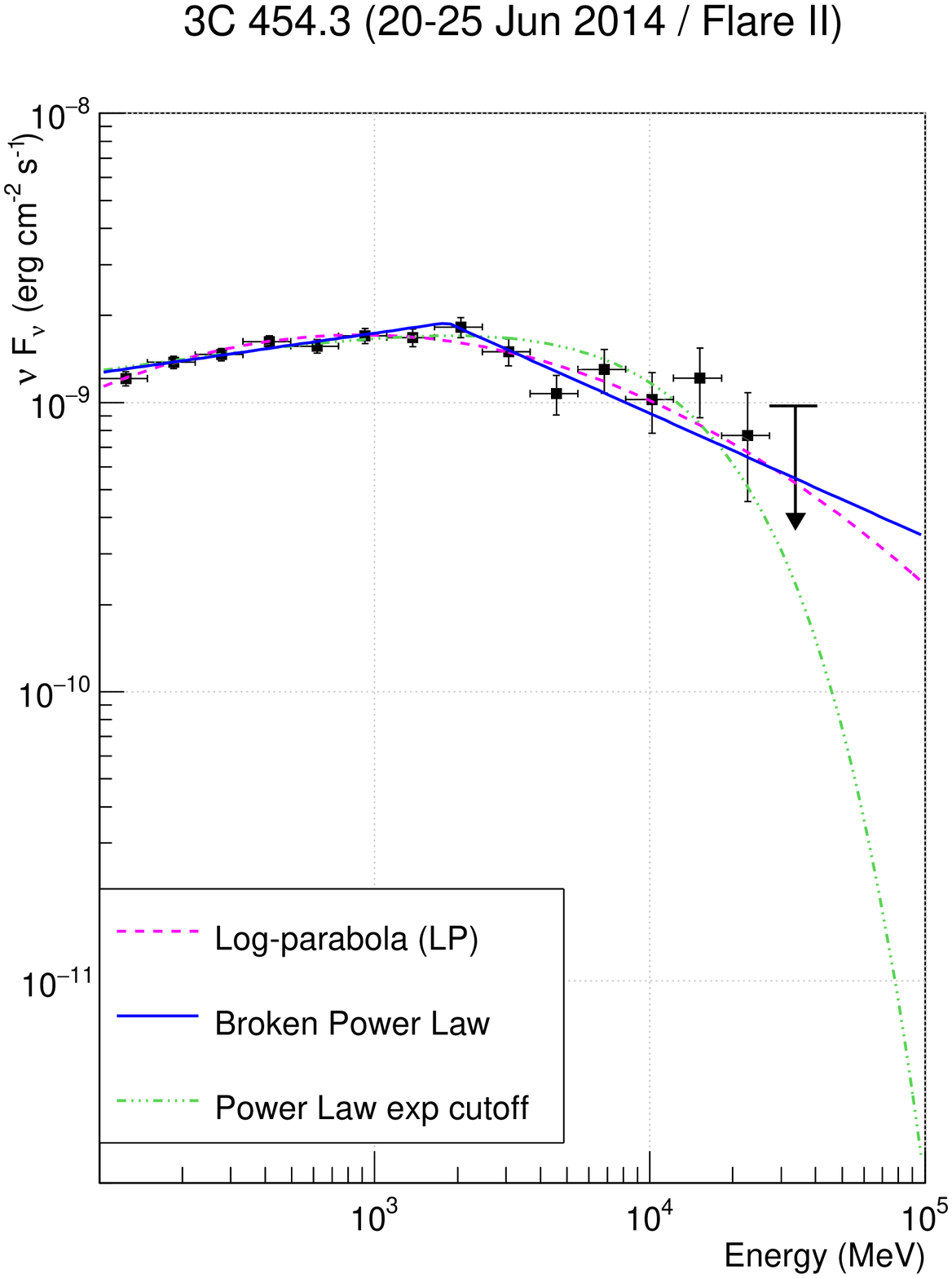}
\caption{Spectral energy distributions of 3C~454.3 above 100 MeV for the two major peaks (Flare I and II) of the flare phase, along with the LP, BPL and PLEC fitted functions. \label{fig:SEDs2}}
\end{figure*}
\begin{figure*}
\includegraphics[width=.32\textwidth]{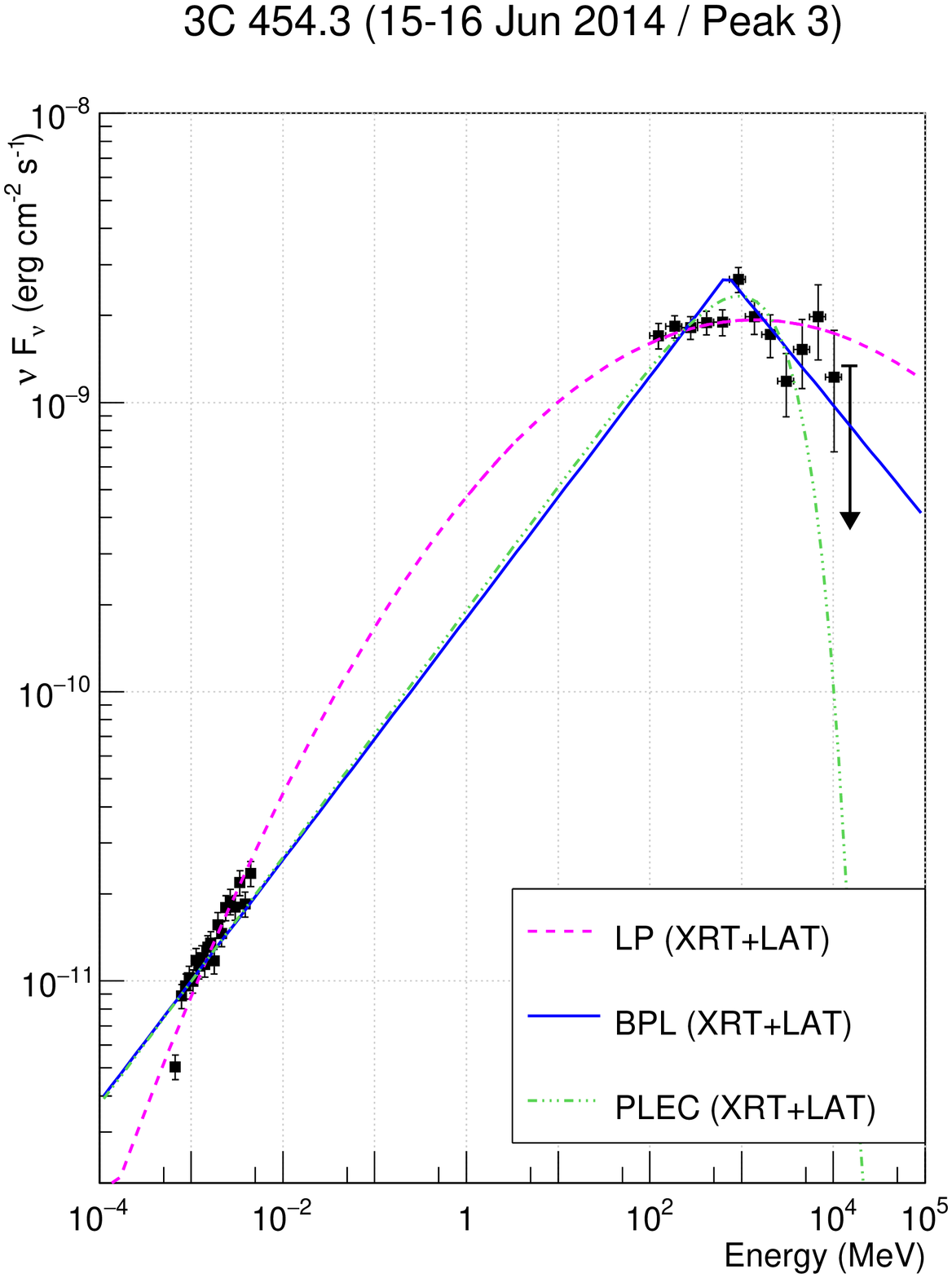}
\includegraphics[width=.32\textwidth]{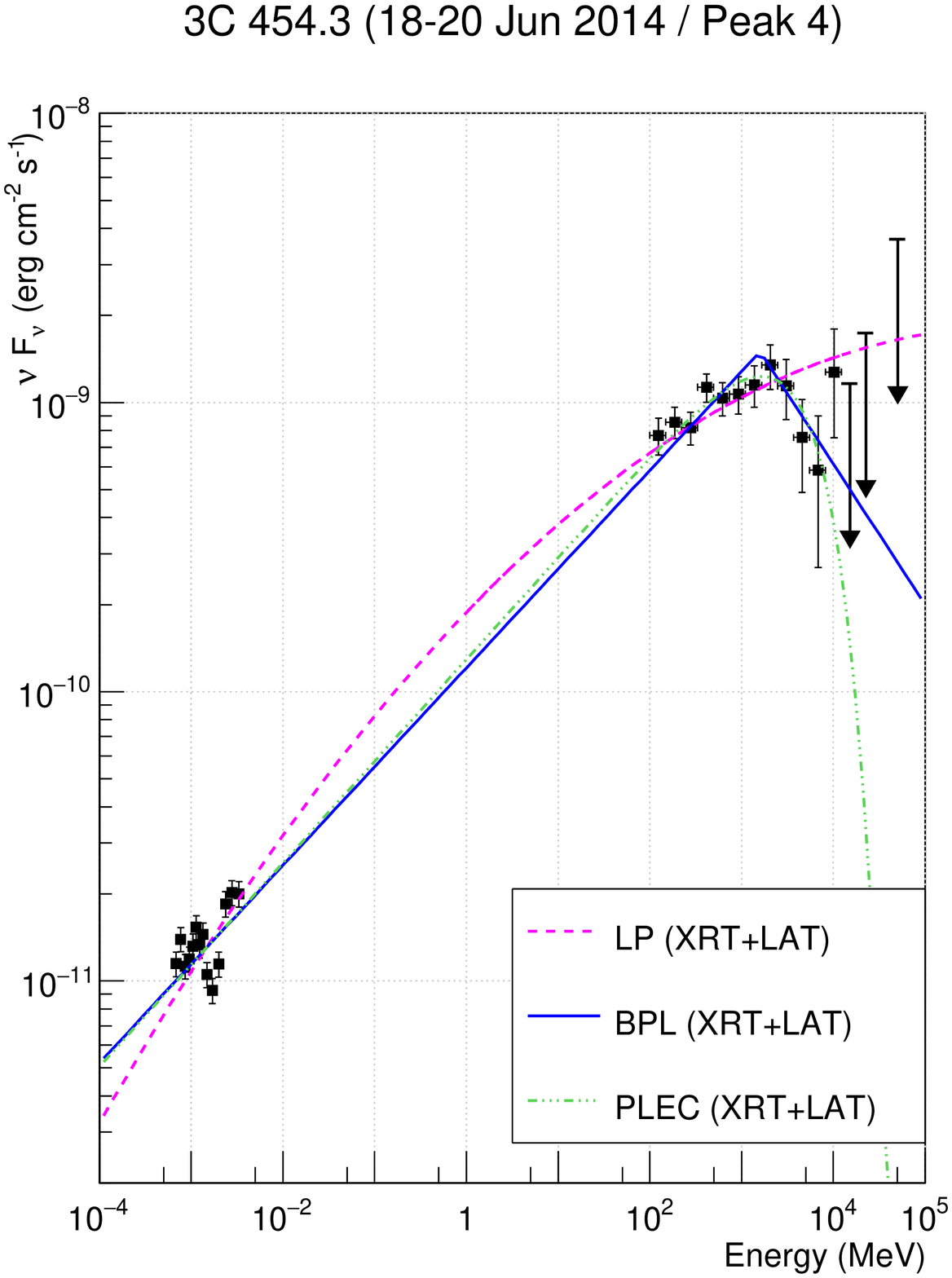}
\includegraphics[width=.32\textwidth]{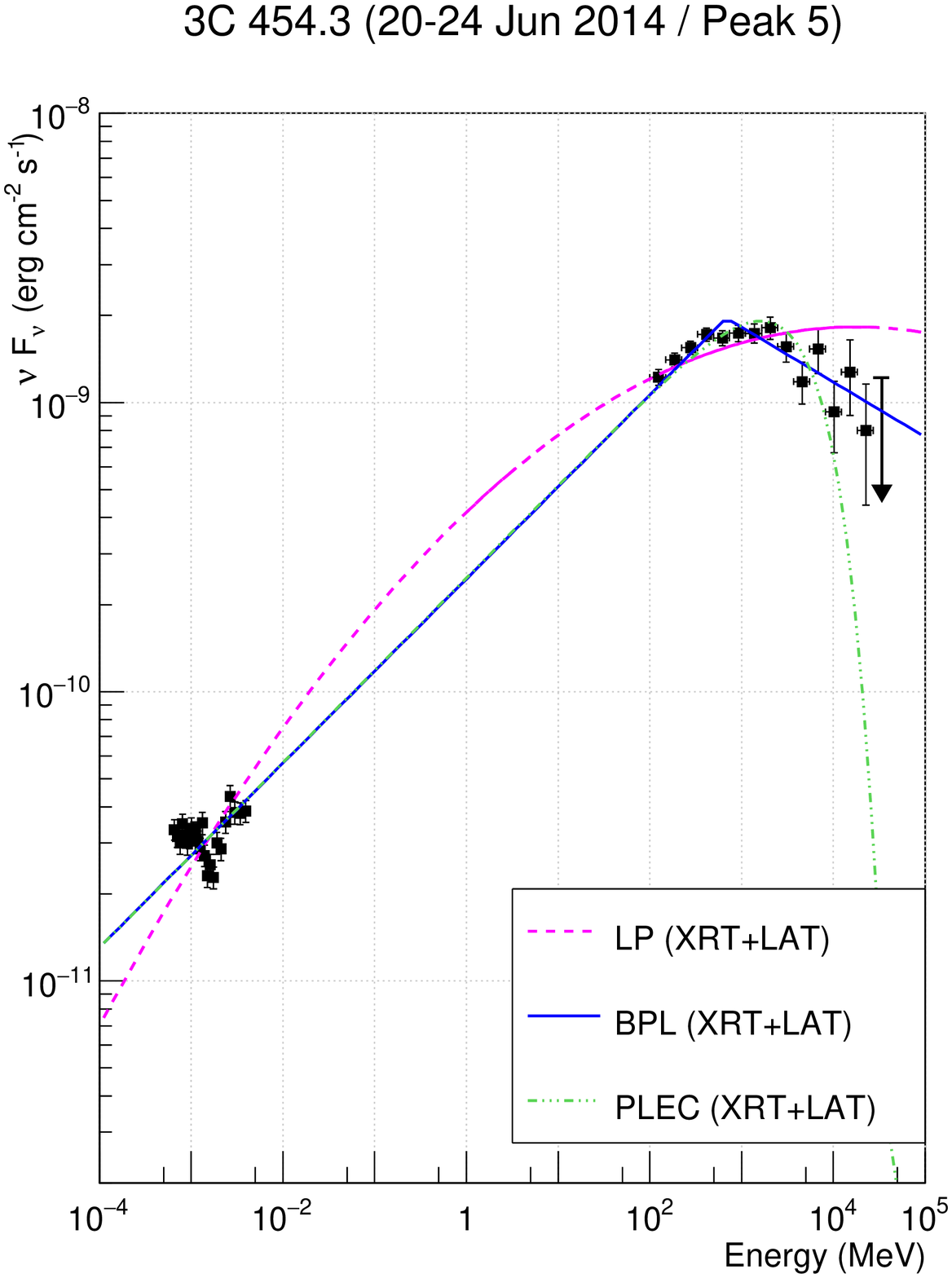}
\caption{Spectral energy distributions of 3C~454.3 in the 1 keV--100 GeV range, with \emph{Swift}-XRT and \emph{Fermi}-LAT data points fitted with LP, BPL and PLEC fitted functions, for Peak 3 (left), Peak 4 (middle) and Peak 5 (right). \label{fig:SEDs3}}
\end{figure*}

Fig. \ref{fig:fast} (left) displays different features concerning the MJD 56816 (``Peak-1'') flare. The top panel displays the LAT exposure rate for 3C~454.3. The bottom panel shows the 3 hr and 6 hr light-curves as well as the function fitted on the 3 hr points (blue) and the function resulting from the fit by the unbinned method (red). To enable a verification of the fit quality, a comparison between the counts of photons ascribed to the source between the data (blue) and the estimate from the fitted function (red) is given in the middle panel. The unbinned method gives $T_r$=4.2$\pm$0.9 ks (1.2$\pm$0.3 hr), $T_f$=15$\pm$3 ks ($4.2 \pm 1.1$ hr) in good agreement with the results of the conventional fit, but with better-defined uncertainties. For the MJD 56827 flare (``Peak 4''), the unbinned method yields $T_r$=1.2$\pm$0.7 ks (0.3$\pm$0.2 hr) and $T_f$=34$\pm$4 ks (9.4$\pm$1.1 hr). $T_r$ in Peak 4 is the shortest timescale reported for this source in the $\gamma$-ray band. Unexpectedly, this flare peaks in flux almost exactly simultaneously with a change in scanning mode. Monte Carlo simulations were performed with the package gtobssim (included in the Fermi Science Tools) assuming a steady source and the actual exposure profile around the time of Peak 4. The results of the unbinned analysis were checked to be robust against a change in scanning mode. This was expected since the method considers only the instantaneous exposure rate. Inspection of the data did not reveal any change in background during this flare that could have biased our results. The 45-GeV photon mentioned in section \ref{subsec:HE_L} was detected at MJD 56827.1233, i.e., 10 ks (2.8 hr) after the flux peak, found at MJD 56827.0078 (Table \ref{tbl-LC_fits}).

\subsection{Time-resolved SEDs}

The 2014 June outburst is less intense than the 2009 December and 2010 November outbursts, and reached a flux level similar to the 2010 April outburst. During these previous flaring activities, evidence of some hardening of the photon spectral index was found. The two-day flare of 2013 September 23--25 had a flux below $F_{100}=4$; however, a harder index $\Gamma \simeq 1.8 \pm 0.1$ was reported in \citet{Pacciani}. The observation and study presented in the previous sections showed that the 2014 May--July outburst also exhibited a dramatic hardening of the photon index. In this section we focus on the SED studies, and report spectral features that characterized the different phases/subphases of the outburst.

We have performed the spectral analysis of the three phases of the 2014 May--July outburst, as well as the quiescent phase that preceded them, and separately the Flare I and Flare II phases (Table \ref{tbl-periods}), in the 0.1--300 GeV energy range, as described in Section \ref{sec:obs_ana}. Spectral models of 3C~454.3, in the likelihood analysis, were represented successively by the four functions defined in Eqs. (\ref{Eq:PL}, \ref{Eq:BPL}, \ref{Eq:LP}, \ref{Eq:PLEC}) indicated in Section 2 (respectively PL, BPL, LP and PLEC). The Log(likelihood) value is returned, corresponding to the best fit model parameters of the sources in the ROI and the source region. The so-obtained fitted functions were plotted with the SED data points, for the LP, BPL and PLEC models respectively. The quality of unbinned fits is given by the Log(likelihood) for each of these three fitting functions, and is reported in Table \ref{tab:fits}.

SED data points were computed in equally spaced logarithmic bins from 100 MeV to the highest energy photon detected (data point or upper limit). The {\em prefactors} (normalization factor) of the three point sources within the ROI, as for the Galactic diffuse and isotropic models, were kept free in the likelihood optimization procedure, whereas all other parameters were fixed to the 3FGL values. When the number of predicted photons associated with 3C 454.3 $N_{\rm pred} \le$ 3 (but $>0$), or the significance of the detection is low ({\em Test Statistic}--TS $<9$), we calculated an upper limit for the corresponding energy bin. This latter computation was performed using the standard {\em UpperLimits} class of the \emph{Python} Fermi Science Tools, by processing outputs of the actual unbinned likelihood analysis, where 3C~454.3 is modeled by a PL.

Results are presented in Fig. \ref{fig:SEDs} and \ref{fig:SEDs2}. The results of the spectral analysis were also used to produce Fig. \ref{fig:Gamma_vs_Flux_periods} (Section \ref{subsec:LCs}). We observe an overall good compatibility between the LP, BPL and PLEC models. Luminosity ($>$ 100 MeV) was also estimated, using Eq. \ref{eq:L}. A progressive hardening with the increase of flux is observed from Phase 1 to 3, at energies $E<E_{\rm break}$, with $\Gamma$=$2.40 \pm 0.01$, $2.27 \pm 0.03$, and $2.04 \pm 0.01$ respectively (for PL models). Hardening of the spectrum is significant during Flare II, with $\Gamma=1.99 \pm 0.02$, compared to $\Gamma=2.08 \pm 0.02$ during Flare I (PL models). To the best of our knowledge, Flare II represents the first extended phase when {\em Fermi}-LAT recorded such a hard spectrum during a major outburst of this source. The effect of this hardening can be seen on the parameters $\Gamma_1$, $\Gamma_2$ and $\alpha$ (Table \ref{tab:fits}). We also fitted the data points with the same functions, but by including the model of absorption by the extragalactic background light (EBL) from \citet{Razzaque} and \citet{Finke}. But none of our results were significantly changed.

We have also modeled an absorption pattern which includes the previously mentioned EBL along with absorption of $\gamma$ rays in the BLR by the two-photon pair production process, as described in \citet{britto}. No absorption feature was found, using a six-line model of the BLR, but the sensitivity of {\em Fermi}-LAT in the range where the absorption could show up ($> 10$ GeV) is too low within the few week duration of the outburst.

\section{Discussion} \label{sec:disc}
The 2014 May--July outburst of 3C~454.3 bore a flaring pattern
typical of this FSRQ, namely a plateau, flare and post-flare phases, which
was also observed during the previous three major $\gamma$-ray outbursts. The flux of 3C~454.3 reached an average value of
$F_{100} = 7.2 \pm 0.2$ during the flaring phase of 2014 June 7--29, which is
similar to the 2010 April flare but much weaker than the giant flare
in 2010 November. The peak flux, $F_{100} = 17.6 \pm 1.9$ was recorded on 2014 June 15 (MJD 56823.5625)
in a 3 hr light-curve. This corresponds to an
isotropic-equivalent $\gamma$-ray luminosity of
$L_\gamma \approx 60 \times 10^{48}$~erg~s$^{-1}$, roughly an order of magnitude lower than
the peak luminosity of the giant flare in 2010 November, but still quite
substantial for a blazar. Therefore the jet opening angle should be
$\theta_j \approx 3^\circ$ in order for the true luminosity
$2(1-\cos\theta_j)L_\gamma$ to be of the order of the accretion disc
luminosity $L_d \approx 6.75\times 10^{46}$~erg~s$^{-1}$ \citep{Bonnoli}. From \emph{Very Long Baseline Interferometry (VLBI)} observation, the jet opening angle is determined to be $0\fdg8\pm0\fdg2$ \citep{2005AJ....130.1418J}.

The 2014 June 7--29 flaring state of 3C~454.3 is characterized by several subflares with distinct peaks. An extremely fast rise of flux
was recorded, on a time scale $T_r\approx 1200$~s, on MJD 56827. This is one of the shortest flux variability times measured for blazars in
the GeV range. A few other subflares also show rapid, on hours time scale, flux rise. These variations could indicate $\gamma$ rays being
emitted from compact regions, namely blobs, in the jet. The radii of these blobs in the comoving frame, $R^\prime \approx \delta ct_v/(1+z)$,
depend on their Doppler factor $\delta$ and the time scale $t_v$ over which their flux vary, e.g. \citet{2008ApJ...686..181F}.
A constraint on $\delta$ can be obtained by requiring the blob to be optically thin to HE
photons against $\gamma\gamma\to e^\pm$ pair production process. In
particular $\gamma$-rays of energy $E_\gamma$ from 3C~454.3 should be
interacting with $2m_e^2c^4 \delta^2/E(1+z)^2 \approx 1.5
(\delta/10)^2 (E_\gamma/10~{\rm GeV})^{-1}$ keV photons at threshold,
if those X-ray photons are produced in the same blob. A more detailed
calculation, following \citet{Gould+1967}, leads to an opacity formula

\begin{equation}
\tau_{\gamma\gamma} (E_\gamma) = \frac{\delta ct_v}{1+z} 
\pi r_0^2 \left[ \frac{m_e^2 c^4 \delta}{(1+z)E_\gamma} \right]^2
\int_{\frac{m_e^2c^4\delta}{(1+z)E_\gamma}}^{\frac{(1+z)E_\gamma}{\delta}}
\frac{n^\prime (\epsilon^\prime)}{\epsilon^{\prime 2}}
\varphi [S_0 (\epsilon^\prime)] d\epsilon^\prime,
\label{ggopt}
\end{equation}

\noindent
for isotropic distribution of photons in the blob, where
$\epsilon^\prime = (1+z)\epsilon/\delta$ is the target photon energy
in the blob frame and

$$n^\prime (\epsilon^\prime) = \frac{(1+z)^2d_L^2}{\delta^4c^3t_v^2}
\,n \left( \frac{\delta\epsilon^\prime}{1+z} \right)$$ 

\noindent
is the target photon spectrum in the blob frame with $n(\epsilon)$
being the observed spectrum. The function $\varphi [S_0 (\epsilon^\prime)]$,
with $S_0 (\epsilon^\prime) = (1+z)\epsilon^\prime E_\gamma/\delta m_e^2c^4$, is defined by
\citet{Gould+1967} and corrected by \citet{Brown+1973}. The condition
$\tau_{\gamma\gamma}(E_\gamma) =1$ can be translated to a lower limit
on $\delta$.

{\em Swift}-XRT observed 3C~454.3 during its 2014 June 7--29 flare.
We analyzed public XRT data, using the procedure described in section \ref{sec:Swift}.
For each of the five data sets we analyzed, detailed SEDs were produced in the 0.6--6 keV range. We determined the values of $\delta$ for the $\gamma$-ray
subflares Peak 3, Peak 4 and Peak 5, respectively, as denoted in
Fig.~\ref{fig:flare_fits_Tr_Tf}. These subflares also contain a 12
GeV (Peak 3), a 45 GeV (Peak 4) and a 39 GeV (Peak 5) photon, detected
with high confidence. To constrain the Doppler factor $\delta$, we fit
the LAT and XRT data together (Fig. \ref{fig:SEDs3}). Since XRT data were aquired for only five short observation periods (each one lasting from $\sim1600$ to $\sim4000$ s), during MJD 56822--56830, we do not have contemporaneous XRT/LAT coverage for each flaring peak. However, the flux level of the five XRT data sets was observed to be almost constant. We therefore assigned the XRT data sets 00031018016, 00031018019 and 00031018020 to Peak 3, Peak 4 and Peak 5, respectively, in order to perform the combined SED fits.
The best fits of our SEDs are found to be with an LP model
during Peak 3, PLEC during Peak 4, and BPL during Peak 5, assuming
the $\gamma$ rays and X-rays originated from the same blob during those
subflares.
Even though the keV--MeV range is modeled as the combination of several radiation production processes, the total fit is not expected to depart significantly from what we fitted in our phenomenological approach, as we can see in the following two references. \citet{2013ApJ...768...54B} modeled the broadband SED of 3C~454.3 using data from the \emph{Fermi}-LAT Bright AGN Sample---LBAS \cite{2010ApJ...716...30A}. They used both a leptonic (SSC+EC) and a hadronic model in their Fig. 4 and Fig. 7, respectively. \citet{Cerruti} performed a leptonic modeling during a relatively quiescent state of 3C~454.3, in 2008 August, and also during the 2010 November giant flare of this source, and showed the SED fits in their Figure 1.

Assuming that the flux variability time is $t_v = T_r =$ 2.1~hr
for Peak 3, 0.3 hr (1200 s, as found by the unbinned method in Section \ref{subsec:FatestVar}) for Peak 4, and 27.8 hr for Peak 5,
we calculated $\delta \gtrsim 19$, 29 and 14, respectively, requiring
the $\gamma\gamma$ opacity for the 12, 45 and 39~GeV photons
to be less than or equal to 1 according to Eq.~(\ref{ggopt}). (We also checked that other fit functions with similar $\chi^2/ndf$ do lead to similar values of $\delta$.) While the
$\delta$ values for Peak 3 and Peak 5 are similar to the ones obtained
for the 2010 November flare, the value $\delta \gtrsim 29$ for Peak 4 is
higher and is particularly interesting. Using an angle $1^\circ.3$
between the jet and our line of sight, inferred as an average value from long-term VLBI \citep{2005AJ....130.1418J}, and
$\delta = [\Gamma_{\rm jet} (1-\beta_{\rm jet}\cos\theta)]^{-1} \simeq 29$
we calculate the jet Lorentz factor $\Gamma_{\rm jet} \simeq 16$.
This is compatible with the previously estimated value of $\Gamma_{\rm jet}=15.6 \pm 2.2$ reported by \citet{2005AJ....130.1418J} and a little below $\Gamma_{\rm jet} \sim 20$ reported in \citet{Sikora}, though compatibility is not necessarily expected, as our study specifically refers to a flare within a short time range, and non-contemporaneous wih the observations of these previous papers.

\begin{table}[t!]
\begin{center}
\caption{Calculated limits of the values of $\delta$, $\beta_{\rm jet}=\sqrt{\Gamma^2_{\rm jet}-1}/\Gamma_{\rm jet}$, $R^\prime$, $\Gamma_{\rm jet}$ and $r$, corresponding to the Peak 3, Peak 4 and Peak 5 subflaring events. \label{tbl-beta_delta_gamma}}
    \begin{tabular}{l|c|c|c|c|c}
      \hline
      Subflaring events  & $\delta$ & $\beta_{\rm jet}$ & $\Gamma_{\rm jet}$ & $R^\prime$ [cm] & $r$ [cm]\\
      \hline
      Peak 3 ($T_r=2.1$ hr) & 19   & 0.995  & 10 & 2.3$\times 10^{15}$ & 2.5$\times 10^{16}$ \\
      Peak 4 ($T_r=1200$ s) & 29   & 0.998  & 16 & 5.6$\times 10^{14}$ & 1.0$\times 10^{16}$ \\
      Peak 5 ($T_r=27.8$ hr) & 14   & 0.991  & 7  & 2.3$\times 10^{16}$ & 1.8$\times 10^{17}$ \\
      \hline  
    \end{tabular}
\end{center}
\end{table}

The distances of the $\gamma$-ray emitting blobs from the central
black hole can be calculated, using the minimum Doppler factor derived
from the $\gamma\gamma$ opacity condition and the resulting minimum
jet Lorentz factor, as $r \simeq 2\Gamma_{\rm jet}^2ct_v/(1+z)$. For the
Peak 3, Peak 4 and Peak 5 data this would imply distances
$r \gtrsim 2.5\times 10^{16}$~cm, $\gtrsim 1.0\times 10^{16}$~cm and
  $\gtrsim 1.8\times 10^{17}$~cm, respectively.
These distances would locate these three blobs in the outer layers of the canonical BLR. The radius of the BLR was estimated to be $R_{BLR} < 10^{18}$ cm, for 3C~454.3 \citep{Bonnoli,2012MNRAS.421.1764S}.  
The values of $\delta$, $\beta_{\rm jet}$, $\Gamma_{\rm jet}$, $R^\prime$
and $r$ that we calculated are reported in Table \ref{tbl-beta_delta_gamma}.
These progressively larger distances have
implications for HE $\gamma$-ray emission. Interestingly
indeed, significant emission above 10 GeV took place during the later
part of the flare, in particular during Peak 3 and later (see
Figure~\ref{fig:flare_fits_Tr_Tf}). In the context of multiple
$\gamma$-ray emitting blobs for different peaks in the LAT light-curve,
blobs which are slower and optically thick to $\gamma\gamma$
pair production would emit little or no HE photons. This
could be the case for Peak 1 and Peak 2. On the other hand, as we have
also derived from the $\gamma\gamma$ opacity condition previously for
Peak~3, Peak~4 and Peak~5, the blobs which are faster and optically
thin to $\gamma\gamma$ pair production would emit significant
HE photons.

We consider the energy spectral index $\alpha$ of the XRT data sets (when expressed in the dimension of $E(dN/dE)$) marginally consistant with the LAT photon index difference $\Gamma_1 - \Gamma_2$ (reported in Table \ref{tab:fits_peaks}), for the respective peaks to which we associated the XRT data sets. Therefore, as shown in \citet{2014ApJ...795...35B}, this condition opens the possibility that the 1--2 GeV break in the SED could be due to internal absorption by X-ray photons. Assigning the respective $E_{\rm break}$ values to $E_\gamma$, and considering the relevant $\Gamma_{\rm XRT}$ indices\footnote{$\Gamma_{\rm XRT}$ is expressed in the dimension of $(dN/dE)$, while $\alpha_{\rm XRT}$ is given in the dimension of $E(dN/dE)$} of Table \ref{tab:3c454-fit}, we calculated the values of both the Doppler and Lorentz factors for Peak 2, Peak 3, Peak 4 and Peak 5, and found $\delta$=10.2, 13.8, 26.2 and 11.0, and $\Gamma_{\rm jet}$=5.2, 7.1, 14.5 and 5.6, respectively. These values are lower than the ones we reported in Table \ref{tbl-beta_delta_gamma}, though the values for Peak 4 and 5 remain compatible. Still, considering the consistancy of the $\alpha \sim \Delta\Gamma$ relation to be only marginal, and based on the work by \citet{Abdo2009}, we are unable to support the idea that the GeV breaks could arise from $\gamma\gamma$ absorption.

Absorption of $\gtrsim 10$ GeV photons, due to $\gamma\gamma\to e^\pm$
pair production process in the BLR with line and continuum photons,
has been suggested as a plausible explanation of spectral break in the
blazar SEDs. While we do find spectral breaks at $\sim$1--2 GeV for
all phases (see Figs.~\ref{fig:SEDs} and \ref{fig:SEDs2}), compatible
with previous outbursts of 3C~454.3, our fits do not improve
significantly when we include absorption using a BLR model with six of
the strongest lines \citep{britto}. This could imply that the
$\gamma$-ray emission region is located close to the outer edge of the
BLR, as is also implied by the constraints on the Doppler factor.\\

\noindent
{\small{\bf Note added:} A paper by \citet{2016MNRAS.458..354C} on the same flare appeared during the reviewing period of our paper. Their results/conclusions are globally consistent with ours, but we consider that the uncertainty on their shorter time scale derived from their binned light-curves is unrealistically small. This could arise due to a lack of convergence in the likelihood analysis and/or due to the use of extremely short time intervals around the time when the data points were measured. Since they used the “Pass 7 REPROCESSED” data representation, their results can be more directly compared to the results of our conference paper \citep{3C454_Pass7} in which we also used “Pass 7 REPROCESSED”.}

\section*{Acknowledgments}

We thank Jean Ballet for the cleaning of data from bright GRBs and solar flares,
and Justin Finke for his meticulous reading of the draft and his valuable comments.

The \textit{Fermi}-LAT Collaboration acknowledges generous ongoing support
from a number of agencies and institutes that have supported both the
development and the operation of the LAT as well as scientific data analysis.
These include the National Aeronautics and Space Administration and the
Department of Energy in the United States, the Commissariat \`a l'Energie Atomique
and the Centre National de la Recherche Scientifique / Institut National de Physique
Nucl\'eaire et de Physique des Particules in France, the Agenzia Spaziale Italiana
and the Istituto Nazionale di Fisica Nucleare in Italy, the Ministry of Education,
Culture, Sports, Science and Technology (MEXT), High Energy Accelerator Research
Organization (KEK) and Japan Aerospace Exploration Agency (JAXA) in Japan, and
the K.~A.~Wallenberg Foundation, the Swedish Research Council and the
Swedish National Space Board in Sweden.
 
Additional support for science analysis during the operations phase is gratefully acknowledged from the Istituto Nazionale di Astrofisica in Italy and the Centre National d'\'Etudes Spatiales in France.

EB acknowledges NASA grants NNX13AO84G and NNX13AF13G.

RJB and SR acknowledge support from the National Research Foundation, South Africa and the South African Gamma-ray Astronomy Programme (SA-GAMMA).\\


\bibliography{brit1104_arXiv2}
\end{document}